\documentclass[letterpaper,10pt]{article}
 \pdfoutput=1 
\usepackage[T1]{fontenc}
\usepackage{parskip}
\usepackage[font={small}]{caption}
\usepackage[margin=2.5cm]{geometry}
\usepackage{times}
\usepackage[affil-it]{authblk}
\usepackage{listings}

\usepackage{booktabs}
\usepackage{makecell}

\date{}

\newenvironment{frontmatter}%
  {}%
  {}

\usepackage{amsmath,amssymb}
\usepackage{graphicx} 
\usepackage{subfig}
\usepackage{epsfig}
\usepackage{algorithm}
\usepackage{soul}
\usepackage{algpseudocode}
\usepackage{float}
                               \usepackage{amsthm}
                               \newtheorem{theorem}{Theorem}[section]

\newtheorem{lem}[theorem]{Lemma}

\usepackage[english]{babel}
\usepackage{cancel}
\newsavebox\CBox

\newcommand\hcancel[2][0.5pt]{%
  \ifmmode\sbox\CBox{$#2$}\else\sbox\CBox{#2}\fi%
  \makebox[0pt][l]{\usebox\CBox}%
  \rule{\wd\CBox}{#1}}

\newcommand{\nc}{\newcommand}
\nc{\mbf}{\mathbf}
\nc{\R}{\Bbb{R}}
\nc{\N}{\mathcal{N}}
\nc{\U}{\Bbb{U}}
\nc{\X}{\Bbb{X}}
\nc{\intr}[1]{\textnormal{int}#1}
\nc{\mpart}[2]{\frac{\partial #1}{\partial #2}}
\nc{\sfrac}[2]{{\textstyle\frac{#1}{#2}}}
\nc{\mprb}[2][]{\ensuremath{({\cal #2}_{#1})}}
\nc{\nser}[3]{#1_{#2},\ldots,#1_{#3}}
\nc{\nserf}[4]{#1(#2_{#3}),\ldots,#1(#2_{#4})}
\nc{\inprod}[2]{\langle#1,#2\rangle}
\nc{\ndthm}{\hspace*{\fill}$\scriptstyle{\Box}$}
\nc{\vndthm}{~\vspace{-\baselineskip}~\ndthm}
\nc{\wideq}[1][=]{\ensuremath{\ \ #1\ \ }}
\nc{\xs}{{\mathscr X}}
\nc{\us}{{\mathscr U}}
\nc{\Seq}{\Bbb{S}}
\nc{\range}[1]{\textnormal{ran}(#1)}
\nc{\domain}[1]{\textnormal{dom}(#1)}
\nc{\fdef}[3]{#1:#2,\quad #3}
\nc{\bmat}[1]{\begingroup\setlength\arraycolsep{8pt}\begin{bmatrix} #1 \end{bmatrix}\endgroup}
\nc{\cmat}[1]{\begingroup\setlength\arraycolsep{8pt} \left ( \begin{matrix} #1 \end{matrix} \right ) \endgroup}
\nc{\rank}{\textnormal{rank}}
\nc{\FF}{\mathcal{F}}
\nc{\PP}{P}
\nc{\OO}{\Omega}
\nc{\Prb}[1]{\textnormal{Pr}\left ( #1 \right )}
\nc{\myP}{\textbf{P}}
\nc{\myA}{\textbf{A}}
\nc{\myQ}{\textbf{Q}}
\nc{\myC}{\textbf{C}}
\nc{\myR}{\textbf{R}}
\nc{\mySigma}{\boldsymbol{\Sigma}}
\nc{\myPsi}{\boldsymbol{\Psi}}
\nc{\myPhi}{\boldsymbol{\Phi}}
\nc{\myPi}{\boldsymbol{\Sigma}}
\nc{\myBeta}{\boldsymbol{\Gamma}}
\nc{\myL}{\textbf{L}}
\nc{\myJ}{\textbf{J}}
\nc{\myI}{\textbf{I}}
\nc{\myK}{\textbf{K}}
\nc{\mys}{{s}}
\nc{\newspace}{\tilde{x}}
\newcommand{\mR}{\mathcal{R}}
\newcommand{\mRb}{\boldsymbol{\mathcal{R}}}
\newcommand{\mQb}{\boldsymbol{\mathcal{Q}}}
\newcommand{\pt}[0]{p_\theta}
\newcommand{\ptp}[0]{p_{\theta'}}
\newcommand{\expected}[0]{\mathop{\mathbb{E}}}
\nc{\numPred}[1]{{M^\text{p}_{#1}}} 
\nc{\numFilt}[1]{{M^\text{f}_{#1}}} 
\nc{\numBack}[1]{{M^\text{b}_{#1}}} 
\nc{\numCorr}[1]{{M^\text{c}_{#1}}} 
\nc{\numJSmooth}[1]{{M^\text{S}_{#1}}}
\nc{\mytrans}{\mathcal{T}}
\nc{\mySRinv}{\Lambda}
\nc{\be}{z}
\nc{\bee}{z^*}

\begin{document}

\begin{frontmatter}

\title{A Variational Expectation-Maximisation Algorithm for \\Learning Jump Markov Linear Systems} 



\author{Mark P. Balenzuela\footnote{Corresponding author: Mark.Balenzuela@uon.edu.au}, Adrian G. Wills, Christopher Renton, and Brett Ninness}

\maketitle

Faculty of Engineering and Built Environment, The University of Newcastle, Callaghan, NSW 2308 Australia



\begin{abstract}                          
Jump Markov linear systems (JMLS) are a useful class which can be used to model processes which exhibit random changes in behavior during operation. 
This paper presents a numerically stable method for learning the parameters of jump Markov linear systems using the expectation-maximisation (EM) approach.
The solution provided herein is a deterministic algorithm, and is \textit{not} a Monte Carlo based technique. 
As a result, simulations show that when compared to alternative approaches, a more likely set of system parameters can be found within a fixed computation time, which better explain the observations of the system.
\end{abstract}
\end{frontmatter}
%
\clearpage
\section{Introduction}
This paper is concerned with parameter estimation for jump Markov linear systems (JMLS). 
These are systems which exhibit stochastic switching of linear behavior and have a wide range of applications, including target tracking \cite{mazor1998interacting}, econometrics \cite{kim1994dynamic}, telecommunications \cite{logothetis1999expectation}, and fault detection and isolation \cite{hashimoto2001sensor}.
%
System identification for switched systems such as JMLS have previously been applied to problems in fault detection \cite{blackmore2007model}, medical applications \cite{ghahramani2000variational}, and identification of unmanned under water vehicles \cite{gil2009beyond}.

Because JMLS can be considered to be a sub-class of hybrid systems, which we consider to be systems with both discrete and continuous state variables, general approaches available to estimate the parameters of hybrid systems can also be applied to the JMLS class.
One common method for learning hybrid systems is to partition the data into disjoint segments \cite{paoletti2007identification,borges2005switching,pekpe2004identification,yildirim2013online,balakrishnan2004inference,bako2009identification,bako2009line}, and perform system identification for each of these segments. 
The major drawback with this approach is the possibility of arbitrarily split data \cite{paoletti2007identification}, and model estimates being corrupted from data which they did not generate.
Similarly, a weighted least squares approach for JMLS identification has been developed previously \cite{chen2011recursive}, which only considers the most likely model for each time step. 

Alternatively, the maximum likelihood (ML) approach features asymptotic normality and efficiency, and doesn't require the prior to be a hyperparameter \cite{ashley2014sequential}.
JMLS is a challenging class to apply ML to, as for a given sequence of system measurements $y_{1:N}$ and parameter vector requiring identification $\theta$, no closed form solutions are available to calculate the exact data likelihood $\pt(y_{1:N})$ \cite{svensson2014identification}, let alone solve the ML problem \cite{blackmore2007model}
\begin{align} \theta = \arg \max_{\theta'} \ptp(y_{1:N}).\end{align}
%
The expectation maximization (EM) algorithm has been used for ML parameter estimation of dynamic systems since the 70s  \cite{dempster1977maximum}.
For a JMLS system, the EM algorithm can be used for estimating the entire parameter set, a subset of parameters, such as the continuous model parameters \cite{gil2009beyond}, or simply just estimating the active model for each time step \cite{logothetis1999expectation}.

The EM algorithm cycles between two stages, the E-step---where state estimates are obtained using a parameter estimate $\theta$, and an M-step---where the parameters estimates $\theta$ are updated for use in the next E-step.
Importantly the EM algorithm does not maximise the data likelihood directly, but instead maximises a simpler function $\mathcal{Q(\theta,\theta')}$ called the $\mathcal{Q}$-function. Where it is guaranteed by Jensen's inequality that increasing the value of this $\mathcal{Q}$-function increases the likelihood of the observed data \cite{gil2009beyond,blackmore2007model,wills2008parameter}.

This $\mathcal{Q}$-function can be compactly be written as a function of so-called expectation of sufficient statistics, which are generated in the previous E-step.
Computing expectation of sufficient statistics requires joint-smoothed state estimates $\ptp(x_{k:k+1},z_{k:k+1}|y_{1:N})$ for $\forall k=1,\dots,N$, which is a hybrid probability distribution where system parameters $\theta'$ and a sequence of $N$ measurements $y_{1:N}$ are used to estimate the continuous system state $x_k\in \R^{n_x}$ and model active indicated by the discrete variable $z_k \in \{ 1,\dots ,m\}$ with correlation to the next time step.
It is unfortunate that generating the joint-smoothed distribution for a JMLS system has exponential computational cost \cite{ashley2014sequential,blackmore2007model,gil2009beyond,balakrishnan2004inference,ghahramani2000variational,blom1988interacting,bergman2000markov,barber2006expectation,helmick1995fixed,kim1994dynamic}, and therefore for the algorithm to be deemed practical, an approximation is required to be made.

One approximation is to approach the system as an entirely nonlinear one, and use a sequential Monte Carlo (SMC) method \cite{ashley2014sequential,yildirim2013online}. 
While SMC is versatile, it is a computationally intensive approach, and cannot guarantee convergence to a solution without an infinite number of particles \cite{ashley2014sequential,blackmore2007model,schon2011system}.
As a result, in general the log-likelihood of the data will not increase locally with each iteration \cite{ashley2014sequential}, which complicates the detection of a stopping criterion.
Additionally, SMC methods also suffer from sampling issues such as degeneracy and impoverishment, degrading the estimation used within the E-step \cite{sarkka2013bayesian}.
Because of these sources of error, simulations using this method have converged to parameter values with discrepancies, or `bias' to the true solution \cite{ashley2014sequential}.

The so-called stochastic approximation (SA) \cite{robbins1951stochastic,delyon1999convergence,andrieu2005stability} has been used within the EM framework to mitigate some of the issues caused by the use of SMC methods. 
Using this approach the $\mathcal{Q}$-function is replaced with one which weights expectation of sufficient statistics from current and previous iterations, and has been used by \cite{lindsten2013efficient,svensson2014identification} to make more efficient use of particles and improve bias, variance, and computational time.

Additionally, SMC methods can be further improved as linear Gaussian EM has an exact closed-form solution \cite{elliott2002line}, which is able to be partially extended to JMLS systems.
SMC EM algorithms that use Rao--Blackwellized approaches \cite{blackmore2007model,gil2009beyond,svensson2014identification} take advantage of these closed-form solutions to the linear component of the problem, resulting in algorithms which are computationally cheaper and have improved local convergent properties.
This approach works by using Monte Carlo techniques to sample a discrete sequence indicating the active model for each time step, turning the estimation problem into a linear time-invariant (LTI) one, which can be solved exactly with linear-Gaussian estimators \cite{kalman1960new,rauch1965maximum,fraser1967new,mayne1966solution}.
The downside to this approach is that only a small subset of the model sequences can be considered, and these model sequences are determined at random.

As an alternative to these SMC approaches, a variational JMLS EM algorithm has been developed previously \cite{ghahramani2000variational}, however the solution uses a unimodal Normal distribution assumption, which degrades the estimate.
In our previous work \cite{generalJMLSpaperBalenzuela}, we approximated mixture distributions with a mixtures of fewer components using a pairwise merging method called Kullback--Leibler reduction (KLR) \cite{runnalls2007kullback}.
This avoided any unimodal Normal approximations, allowed the user to trade accuracy for computational cost, and could generate the exact solution where computationally feasible.

In this paper, we develop an EM algorithm, built on the foundation of this previous work, \cite{generalJMLSpaperBalenzuela}, which is deterministic in nature and allows the user to trade computational time for accuracy.
As online EM algorithms commonly approximate the the E-step with the simpler filtered distribution \cite{ozkan2012online,balakrishnan2004inference}, we restrict this paper to the development of accurate offline implementations of the EM algorithm which require smoothed distributions.

\textbf{{\em The contribution}} of this paper is: 
\begin{enumerate}
\item A JMLS joint Two-Filter smoother, which is capable of generating the exact solution when enough computational power is available.
\item A compact closed-form algorithm which is an exact solution to the M-step for the distribution provided by (1), including the calculation of the cross-covariance term $\mathbf{S}(z)$, which we introduce later. 
\item Numerically stable implementation of the above. Note that (1), (2) and (3) results in a fully deterministic numerically stable EM algorithm.
\item Some convergence analysis on the proposed algorithm, with regard to initialisation parameters.
\end{enumerate}

The remainder of this paper is organised as follows. Section~\ref{sec:problem-formulation} provides technical detail about the problem, 
Section~\ref{sec:theProposedAlgorithm3} provides details and instructions on implementing the proposed method, and Section~\ref{sec:EMconsider4} discusses considerations and complications for parameter estimation of JMLS systems.
Section~\ref{sec:simulations5} demonstrates the effectiveness of the proposed method, by comparing it to alternate approaches.
Finally, we provide concluding remarks in Section~\ref{sec:conclusion}.

\clearpage
\section{Problem formulation}\label{sec:problem-formulation}
This paper is directed towards efficiently solving the expectation-maximisation problem for a JMLS system defined by
\begin{subequations}
\label{eq:JMLSdef1}
\begin{align}
  X_{k+1} &= \mathbf{A}(Z_k)X_k + \mathbf{B}(Z_k)u_k + V_k(Z_k),\\
  Y_k &= \mathbf{C}(Z_k)X_k + \mathbf{D}(Z_k)u_k + E_k(Z_k),
\end{align}
where the matrices $\textbf{A}(Z_k)$, $\textbf{B}(Z_k)$, $\textbf{C}(Z_k)$ and $\textbf{D}(Z_k)$ are model parameters with suitable dimensions, indexed by the discrete random variable $Z_k$, which operates according to a Markov chain governed by transition probabilities within the matrix $\mathbf{T}$. i.e., the probability of transitioning from the $i$-th model to the $j$-th model is given by element \mbox{$\mathbf{T}(j,i)=\mathbb{P}(Z_{k+1}=j|Z_k=i)$}.
Additionally, $V_k(Z_k)$ and $E_k(Z_k)$ are random variables from the Gaussian white noise process
\begin{align}
  \label{eq:autosam:3}
  \begin{bmatrix} V_k(Z_k)\\E_k(Z_k) \end{bmatrix} &\sim \mathcal{N}\left(\vec{0},\begin{bmatrix} \mathbf{Q}(Z_k) &\mathbf{S}(Z_k) \\ \mathbf{S}^T(Z_k) & \mathbf{R}(Z_k)\end{bmatrix}\right),
\end{align}
\end{subequations}
parameterised by the covariance matrices $\mathbf{Q}(Z_k)$ and $\mathbf{R}(Z_k)$, and cross-covariance matrix $\mathbf{S}(Z_k)$.

In part, identification of this system is completed by providing a closed-form expression for the parameter set $\theta$ defining the JMLS system which maximizes the $\mathcal{Q}$-function 
\begin{small}
\begin{align}
\label{eq:fhue3}
\mathcal{Q}(\theta,\theta') \triangleq  \sum_{z_{1:N+1}}\int \ln&\left(\pt \left(x_{1:N+1},z_{1:N+1},y_{1:N}\right)\right)  \ptp(x_{1:N+1},z_{1:N+1}|y_{1:N}) \, dx_{1:N+1}.
\end{align}
\end{small}
Notice that the joint-smoothed probability distribution \\ $\ptp(x_{1:N+1},z_{1:N+1}|y_{1:N})$ is not actually required, as 
the joint probability term $\pt \left(x_{1:N+1},z_{1:N+1},y_{1:N}\right)$ is comprised of a sum of functions each over a smaller region of the joint state space, and therefore \eqref{eq:fhue3} only requires the partial joint smoothed distribution $\ptp(x_{k:k+1},z_{k:k+1}|y_{1:N})$ for $k=1,\dots,N$, since
\begin{subequations}
\begin{align}
\label{eq:joint45454565}
\ln&\left(\pt \left(x_{1:N+1},z_{1:N+1},y_{1:N}\right)\right) = \sum_{k=1}^{N} \ln\left(  \N \left( \begin{bmatrix} y_k\\ x_{k+1}   \end{bmatrix} \bigg| \boldsymbol{\Gamma}(z_k)\begin{bmatrix} x_{k} \\ u_{k} \end{bmatrix},\boldsymbol{\Pi}(z_k) \right)\right) +\ln\left(\frac{\mathbf{T}(z_{k+1},z_k)}{\sum_{\ell=1}^{m}\mathbf{T}(\ell,z_k)}\right),
\end{align}
where
\begin{align}
& \boldsymbol{\Gamma}(z_k) =  \begin{bmatrix}\mathbf{C}(z_k) & \mathbf{D}(z_k) \\ \mathbf{A}(z_k) & \mathbf{B}(z_k)  \end{bmatrix}, \\
&\boldsymbol{\Pi}(z_k) = \begin{bmatrix} \mathbf{R}(z_k)& \mathbf{S}^T(z_k) \\ \mathbf{S}(z_k) & \mathbf{Q}(z_k) \end{bmatrix}.
\end{align} 
\end{subequations}

Notice the division by the sum of the transition components in the last term of \eqref{eq:joint45454565} is included to build in the total law of probability to the closed-form expression, this is a similar approach to that taken in \cite{blackmore2007model}.

\textbf{Problem Statement:} 
Given a finite sequence of observations $y_{1:N}$ and a finite sequence of control inputs $u_{1:N}$, where
\begin{align}
u_{1:N} = \{ u_1,\dots,u_{N} \}, \quad y_{1:N} = \{ y_1,\dots , y_N\},
\end{align}
with $u_k \in \mathbb{R}^{n_u}$ and $y_k \in \mathbb{R}^{n_y}$,
determine the model parameters $\theta$ that maximise the likelihood $\pt(y_{1:N})$.
Where the $\theta$ vector fully parameterises the transition matrix for the Markov chain $\mathbf{T}$ as well as the parameters for each of the available continuous systems $\{\boldsymbol{\Pi}(z),\boldsymbol{\Gamma}(z) \}_{z=1}^m$.

\clearpage
\section{The JMLS EM Algorithm}
\label{sec:theProposedAlgorithm3}
This section details the operation of the proposed EM algorithm for identifying a JMLS system.

\subsection{Transforming the JMLS system}
In order to use the smoothing solution provided in \cite{generalJMLSpaperBalenzuela}, it is necessary to transform the system provided by the initial guess or previous M-step into an equivalent system with the form
\begin{subequations}
\begin{align}
X_{k+1} &= \mathbf{A}_k(Z_k) X_k + \mathbf{B}_k(Z_k)\bar{u}_k +V_k(Z_k),\\
Y_k &= \mathbf{C}_k(Z_k) X_k + \mathbf{D}_k(Z_k)\bar{u}_k + E_k(Z_k),
\end{align}
where,
\begin{align}
V_k(Z_k) &\sim \mathcal{N}(0,\mathbf{Q}_k(Z_k)),\\
E_k(Z_k) &\sim \mathcal{N}(0,\mathbf{R}_k(Z_k)).
\end{align}
\end{subequations}
This new system description has a different noise structure, which does not have correlation between process and measurement noise channels, opposed to the previous structure \eqref{eq:JMLSdef1}.
Also note that the system now has time varying parameters within the E-step, which is perfectly allowed in the joint smoothing solution, but also uses a different input vector $\bar{u}_k$.
The new input and parameters for the transformed system can be found by the relations
\begin{subequations}
\label{eq:conversionsystem378}
\begin{align}
&\mathbf{A}_k(z_k) = \mathbf{A}(z_k)-\mySRinv_k(z_k) \mathbf{C}(z_k),\\
&\mathbf{B}_k(z_k) = \begin{bmatrix}\mathbf{B}(z_k) -\mySRinv_k(z_k)\mathbf{D}(z_k) &\ \mySRinv_k(z_k) \end{bmatrix},\\
&\mathbf{C}_k(z_k) =\mathbf{C}(z_k),\\
&\mathbf{D}_k(z_k) = \begin{bmatrix} \mathbf{D}(z_k) & \ \mathbf{0}_{n_y} \end{bmatrix},\\
%
&\mySRinv_k(z_k) = H^T(z_k)(\mathbf{R}^{1/2}_k(z_k))^{-T},\\ 
&\begin{bmatrix} \mathbf{R}^{1/2}_k(z_k) & H(z_k) \\ \mathbf{0} & \mathbf{Q}^{1/2}_k(z_k) \end{bmatrix} = \mathbf{\Pi}^{1/2}(z_k),\\
&\bar{u}_k = \begin{bmatrix}u_k\\y_k\end{bmatrix}.
\end{align}
\end{subequations}
Where $A^{1/2}$ denotes the upper Cholesky factor of A, and therefore \mbox{$(A^{1/2})^TA^{1/2}=A$}, where $A$ is an upper triangular (UT) matrix.

\subsection{Numerically stable joint smoother}
In this section we build on our previous contribution \cite{generalJMLSpaperBalenzuela}, and will assume the statistics of the JMLS forwards filter and backwards information filter (BIF) are readily available for $\forall k=1,\dots,N$.
We now provide a Lemma detailing the numerically stable calculation of the joint smoothed distribution for a JMLS system.
All proofs are provided in the Appendix. 
\begin{lem}
\label{lem:jointsmoothedmain43}
\begin{small}
For a given forward filtered distribution
  \begin{align}
    \label{eq:MCHA6100 Notes on JMLS:10}
    p(x_k,z_k \mid y_{1:k}) &= \sum_{i=1}^{\numFilt{k}} w^i_{k | k}(z_k) \, \mathcal{N} \left (x_k\, | \, \mu^i_{k
                              \mid k}(z_k), \, \myP^i_{k \mid k}(z_k)
                              \right ),
  \end{align}
  and backwards filtered likelihood
    \begin{align}
    \label{eq:MCHA6100 Notes on JMLS:29}
    p&(y_{k+1:N} | x_{k+1},z_{k+1}) = \sum_{i=1}^{\numCorr{k+1}} \mathcal{L} \left ( x_{k+1} \, \big|\, \bar{r}_{k+1}^i(z_{k+1}),\, \bar{s}_{k+1}^i(z_{k+1}),\, \bar{\myL}_{k+1}^i(z_{k+1}) \right ),
    \end{align}
where $\mathcal{L}( x \, |\, r,\, s,\, \myL )$ denotes a quadratic likelihood function which is defined as
    \begin{align}
  \mathcal{L} &\left( x \, |\, r,\, s,\, \myL\right) \triangleq e^{-\frac{1}{2}\left(r + 2x^T s + x^T \myL x\right )}.
\end{align}
Then using the transformed time varying hybrid transition model
\begin{align}
  p&(x_{k+1},z_{k+1}|x_k,z_k)  = \mathbf{T}(z_{k+1},z_k))\mathcal{N}(x_{k+1}|\mathbf{A}_k(z_k)x_k+b_k(z_k),\mathbf{Q}_k(z_k)),
  \end{align}
the joint smoothed distribution can be formed.
The joint smoothed distribution
  \begin{align}
  p&(x_{k+1},z_{k+1},x_k,z_k \mid y_{1:N}) =  \sum_{j=1}^{\numJSmooth{k}} w^j_{k:k+1 | N}(z_{k+1},z_k)  \mathcal{N} \left ( \begin{bmatrix}x_k \\ x_{k+1} \end{bmatrix}\, \bigg| \, \mu^j_{k:k+1
                              \mid N}(z_{k+1},z_k), \, \myP^j_{k:k+1 \mid N}(z_{k+1},z_k)
                              \right ),
  \end{align}
can be computed by calculating the defining statistics using the following equations.
It is highly recommended that these statistics be calculated suing a log-weight implementation of
\begin{subequations}
\begin{align}
&\numJSmooth{k}  = \numFilt{k}\cdot \numCorr{k+1},\\
j &= \numFilt{k} (\ell - 1) + i,\\
{w}&_{k:k+1 | N}^j(z_k,z_{k+1})  \\
&= \frac{\tilde{w}_{k:k+1 \mid N}^j(z_k,z_{k+1})}{\sum_{z_{k+1}=1}^{m}\sum_{z_k=1}^{m} \sum_{p=1}^{\numJSmooth{k}} \tilde{w}_{k:k+1 \mid N}^p(z_k,z_{k+1})},\\
\tilde{w}&_{k:k+1 \mid N}^j(z_k,z_{k+1}) = e^{\frac{1}{2}\beta^j(z_k,z_{k+1})} ,\\ 
        \beta&^j(z_k,z_{k+1}) \nonumber \\
        &= (\mu^j_{k:k+1|N}(z_k,z_{k+1}))^T(\myP_{k:k+1|N}^j(z_k,z_{k+1}))^{-1}\nonumber \\ 
        & \quad \cdot \mu^j_{k:k+1|N}(z_k,z_{k+1}) + 2\ln(w^i_{k|k}(z_k)) + 2\ln(\mathbf{T}(z_{k+1},z_k)), \nonumber \\
    &\quad -  (\mu^i_{k:k+1 \mid k}(z_k))^T (\myP^i_{k:k+1 | k}(z_k))^{-1} \mu^i_{k:k+1 | k}(z_k)  \nonumber \\
    &\quad  -\bar{r}^\ell_{k+1}(z_{k+1}) +\ln|\myP_{k:k+1|N}^j(z_k,z_{k+1})| - \ln|\myP_{k:k+1|k}^i(z_k)|  \nonumber \\
     \mu&^j_{k:k+1 | N}(z_k,z_{k+1}) = \myP^j_{k:k+1|N}(z_k,z_{k+1}) \nonumber \\
& \cdot \left(  (\mathbf{P}^i_{k:k+1 | k}(z_k))^{-1} \mu^i_{k:k+1|k}(z_k) -\gamma^\ell_{k+1}(z_{k+1})  \right), \\
\gamma&^\ell_{k+1}(z_{k+1}) = \begin{bmatrix} \vec{0}_{n_x } \\ \bar{s}^\ell_{k+1}(z_{k+1})\end{bmatrix}, \\
\mu&^i_{k:k+1|k} (z_k) = \begin{bmatrix} \mu_{k|k}^i(z_k) \\ \mathbf{A}_k(z_k)\mu^i_{k|k}(z_k) + b_k(z_k)\end{bmatrix},
 \end{align}
\end{subequations}
where $\vec{0}_{n}$ denotes a column vector of zeros with length $n$.
Forming the square-root factor of the joint-smoothed covariance matrix requires a Q-less QR decomposition as
 \begin{subequations}
 \begin{align}
& (\mathbf{P}^j_{k:k+1|N}(z_k,z_{k+1}))^{1/2} = \mR_{22}, \\
&\begin{bmatrix} \mR_{11} & \mR_{12} \\0 & \mR_{22} \end{bmatrix} =\mQb\begin{bmatrix} \mathbf{I}_{{2 n_x}} & \mathbf{0}_{2n_x} \\ \mathbf{J}_k^j(z_k,z_{k+1}) & (\mathbf{P}^i_{k:k+1|k}(z_k))^{1/2}  \end{bmatrix},\\
& (\mathbf{P}_{k:k+1|k}^i(z_k))^{1/2} = \begin{bmatrix} (\mathbf{P}^i_{k|k}(z_k))^{1/2} & (\mathbf{P}^i_{k|k}(z_k))^{1/2}\mathbf{A}_k^T(z_k) \\ \mathbf{0}_{{n_x}} & \mathbf{Q}^{1/2}_k(z_k)\end{bmatrix},\\
& \mathbf{J}_k^j(z_{k},z_{k+1}) = \begin{bmatrix} \mathbf{0}_{{n_x}}& (\mathbf{P}^i_{k|k}(z_k))^{1/2}\mathbf{A}_k^T(z_k)((\bar{\mathbf{L}}_{k+1}^\ell(z_{k+1}))^{1/2})^T \\ \mathbf{0}_{{n_x}} & \mathbf{Q}^{1/2}_k(z_k)((\bar{\mathbf{L}}_{k+1}^\ell(z_{k+1}))^{1/2})^T\end{bmatrix}.
 \end{align}
 \end{subequations}
Where $ \mathbf{0}_{{n}}$ denotes the zeros matrix with dimension \mbox{$n \times n$}, and $\mathbf{I}_{{n}}$ is the identity matrix of size $n\times n$. 
%
\end{small}
\end{lem}

\subsection{The JMLS EM algorithm}
After computing the statistics joint-smoothed distribution using Lemma~\ref{lem:jointsmoothedmain43}, the expectation of sufficient statistics can be generated to complete the E-step.
The explicit equations with full notation for completing this is cumbersome, and as such we define the following shorthand for the expectation operator over a Gaussian component in the joint smoothed mixture
\begin{small}
\begin{align}
&\expected_{k}^{z_k,z_{k+1},\ell}\left[f(X_k,X_{k+1}) \right] \triangleq \int f(x_k,x_{k+1})  \mathcal{N}\left(\begin{bmatrix}x_k \\ x_{k+1} \end{bmatrix}\bigg|\mu_{k:k+1|N}^\ell  (z_k,z_{k+1}), \mathbf{P}_{k:k+1|N}^\ell (z_k,z_{k+1}) \right) \, dx_{k:k+1}.
\end{align}
\end{small}
%
%
Where the expectations required for calculating the expectation of sufficient statistics have closed-form solutions provided by the following Lemma.
\begin{small}
\begin{lem}\label{lem:expectationNormal}
Given a joint Normal distribution parameterised by $\mu_{k:k+1}$ and $\mathbf{P}_{k:k+1}$, written as
\begin{align}
\mathcal{N}\left(\begin{bmatrix}x_k \\ x_{k+1} \end{bmatrix}\bigg|\mu_{k:k+1} , \mathbf{P}_{k:k+1}) \right),
\end{align}
then the expectations of functions can be computed as
\begin{align}
\expected_{k}&\left[f(X_k,X_{k+1}) \right] = \int f(x_k,x_{k+1})  \mathcal{N}\left(\begin{bmatrix}x_k \\ x_{k+1} \end{bmatrix}\bigg|\mu_{k:k+1}, \mathbf{P}_{k:k+1} \right) \, dx_{k:k+1}.
\end{align}
The equations for the required expectations over a Gaussian distribution are therefore
\begin{subequations}
\begin{align}
\expected_k \left[   \begin{bmatrix} X_k\\u_k \end{bmatrix}\begin{bmatrix} X_k\\u_k \end{bmatrix}^T \right] &=
\begin{bmatrix} E_{11}& \mu_k u_k^T \\
u_k \mu_{k}^T & u_ku_k^T\end{bmatrix},
\\
\expected_k \left[ \begin{bmatrix} Y_k \\ X_{k+1} \end{bmatrix} \begin{bmatrix} Y_k \\ X_{k+1} \end{bmatrix}^T \right] &=
\begin{bmatrix} y_ky_k^T & y_k\mu^T_{k+1} \\
 \mu_{k+1} y_k^T &E_{22} \end{bmatrix},
\\
\expected_k  \left[  \begin{bmatrix} Y_k \\ X_{k+1} \end{bmatrix}   \begin{bmatrix} X_k\\u_k \end{bmatrix}^T\right]  &=
 \begin{bmatrix} 
y_k \mu_{k} & y_ku_k^T  \\
E_{21} & \mu_{k+1}u_k^T\end{bmatrix},
\end{align}
where
\begin{align}
\begin{bmatrix} \mu_{k} \\ \mu_{k+1}\end{bmatrix} &= \mu_{k:k+1}, \ 
\begin{bmatrix}E_{11} & E_{21}^T \\ E_{21} & E_{22}\end{bmatrix}=  \mathbf{P}_{k:k+1}+\mu_{k:k+1}\mu_{k:k+1}^T.
\end{align}
\end{subequations}
\end{lem}
\end{small}
We now provide the closed-form solution for optimal model parameter set $\theta$ according to the maximum of the $\mathcal{Q}$-function, which utilise the provided expectations.
\begin{small}
\begin{lem}
\label{lem:JMLS_EM_soln345}
Given the joint-smoothed distribution \\ $p(x_{k+1},z_{k+1},x_k,z_k|y_{1:N}) \quad \forall k=1,\dots,N$ provided by Lemma~\ref{lem:jointsmoothedmain43}, the optimal parameter set according to a maximum of the $\mathcal{Q}$-function can be found using
\begin{subequations}
\label{eq:eight545}
\begin{align}
&p(x_1,z_1) = p(x_1,z_1|y_{1:N}), \\
%
%
&\boldsymbol{\Pi}(\be) = \frac{1}{c_m(\be)} \Big( \boldsymbol{\Phi}(\be) - \boldsymbol{\Gamma} (\be) \boldsymbol{\Psi}^T(\be)  -\boldsymbol{\Psi}(\be) \boldsymbol{\Gamma}^T (\be)  \nonumber \\
& \quad \quad \quad + \boldsymbol{\Gamma} (\be) \boldsymbol{\Sigma} (\be) \boldsymbol{\Gamma}^T (\be) \Big)  \label{eq:coveEMest22}, \\
%
&\boldsymbol{\Gamma}(\be) = \boldsymbol{\Psi}(\be) \boldsymbol{\Sigma}^{-1}(\be) \label{eq:gammaEMest22}, \\
&\mathbf{T}(\bee,\be) = \frac{\sum_{k=1}^{N} \sum_{j=1}^{\numJSmooth{k}} w^j_{k:k+1|N}(\be,\bee)}{ c_m(\be) } \label{eq:TEMest22} ,\\
&c_m(\be) = \sum_{k=1}^{N} \sum_{j=1}^{m} \sum_{\ell=1}^{\numJSmooth{k}} w^\ell_{k:k+1|N}(\be,j), \\
&\boldsymbol{\Sigma}(\be) = \sum_{k=1}^{N}\sum_{j=1}^{m} \sum_{\ell=1}^{\numJSmooth{k}} w^\ell_{k:k+1|N}(\be,j)  \expected_{k}^{\be,j,\ell} \left[   \begin{bmatrix} X_k\\u_k \end{bmatrix}\begin{bmatrix} X_k\\u_k \end{bmatrix}^T \right] ,
\\
&\boldsymbol{\Phi}(\be)= \sum_{k=1}^{N}  \sum_{j=1}^{m}  \sum_{\ell=1}^{\numJSmooth{k}}w_{k:k+1|N}^\ell (\be,j)   \expected_{k}^{\be,j,\ell} \left[ \begin{bmatrix} Y_k \\ X_{k+1} \end{bmatrix} \begin{bmatrix}  Y_k \\  X_{k+1}\end{bmatrix}^T \right] ,
\\
&\boldsymbol{\Psi}(\be) =\sum_{k=1}^{N} \sum_{j=1}^{m}    \sum_{\ell=1}^{\numJSmooth{k}}w_{k:k+1|N}^\ell (\be,j)  \expected_{k}^{\be,j,\ell} \left[  \begin{bmatrix} Y_k \\ X_{k+1} \end{bmatrix}   \begin{bmatrix} X_k\\u_k \end{bmatrix}^T\right].
\end{align} 
\end{subequations}

The expectation of sufficient statistics $\boldsymbol{\Sigma}(z)$, $\boldsymbol{\Phi}(z)$, and $\boldsymbol{\Psi}(z)$ require many expectations to be calculated over a smaller region of the state-space which can be computed according to Lemma~\ref{lem:expectationNormal}.

Note that if only a reduced set of parameters is to be estimated, \eqref{eq:coveEMest22} doesn't require \eqref{eq:gammaEMest22} to be applied, or vice versa.
Additionally, \eqref{eq:TEMest22} doesn't need to be applied if the transition probabilities are known.
Finally, estimation of each of the models is separated in the $\mathcal{Q}$-function, so not all models are required to have a common unknown parameter set.
In fact, it isn't required for each model to have parameters estimated.
\end{lem}
\end{small}

The equations within \eqref{eq:eight545} are similar to those presented in \cite{svensson2014identification}, but differ as these equations also generate a cross-covariance term $\mathbf{S}(z)$.
This term is particularly important if the estimator is to be used on a real-world system, which switches infrequently, as the cross-covariance $\mathbf{S}(z)$ captures characteristics from integrated sampling of ADCs and discretisation of the continuous system.
Additionally, unlike \cite{svensson2014identification}, we do not require such a large matrix containing expectation of sufficient statistics to be stored, or require a seperate Rauch--Tung--Striebel (RTS) smoother to generate them.
\subsection{Numerically stable EM implementation}
In this section, we document a numerical stable implementation of the proposed EM algorithm.
We begin with the following Lemma which provides a numerically stable formula for calculating the square-root factor of a combined expectation term.
\begin{small}
\begin{lem}
\label{lem:sqrtmassExpect45768}
Given the joint Normal distribution parameterised by $\mu_{k:k+1}$ and $\mathbf{P}^{1/2}_{k:k+1}$, written as
$$\mathcal{N}\left(\begin{bmatrix}x_k\\x_{k+1}\end{bmatrix}\bigg| \mu_{k:k+1},\mathbf{P}_{k:k+1}\right).$$
The square-root factor of the combined expectation over this distribution has the form
\begin{align}
\expected_{k} \left[\begin{bmatrix}X_k\\X_{k+1}\\u_k\\Y_k\end{bmatrix}\begin{bmatrix}X_k\\X_{k+1}\\u_k\\Y_k\end{bmatrix}^T\right]^{1/2} = \begin{bmatrix} M_{11}&M_{12}\\0 & M_{22}\end{bmatrix},
\end{align}
where 
\begin{subequations}
\begin{align}
M_{22} &= \sqrt{1-\lambda^T\lambda} \begin{bmatrix} u_k^T & y_k^T\end{bmatrix},\\
M_{12} &= \lambda \begin{bmatrix} u_k^T & y_k^T\end{bmatrix},\\
\lambda &= M_{11}^{-T}\mu_{k:k+1},\\
M_{11} &= \mRb= \mQb \begin{bmatrix} \mathbf{P}_{k:k+1}^{1/2} \\ \mu_{k:k+1}^T\end{bmatrix}.
\end{align}
\end{subequations}
\end{lem}
\end{small}
Using these instructions for generating the square-root factor of the combined expectation, we now provide now instructions for completing the numerical stable implementation of the EM algorithm in the following Lemma.
%
\begin{small}
\begin{lem}
\label{sec:M-stepqrt346}
Given the joint-smoothed distribution \\ $p(x_{k+1},z_{k+1},x_k,z_k|y_{1:N}) \quad \forall k=1,\dots,N$ provided by Lemma~\ref{lem:jointsmoothedmain43}, the optimal numerically stable parameter set according to the maximum of the $\mathcal{Q}$-function can be found using

\begin{small}
\begin{subequations}
\begin{align}
&\boldsymbol{\Pi}^{{1/2}}(\be) = \frac{1}{\sqrt{c_m(\be)}} \mathbf{M}^{{1/2}}(z)  
\begin{bmatrix} -\mathbf{C}^T(\be) & -\mathbf{A}^T(\be) \\ \mathbf{0}_{n_x\times n_y} & \mathbf{I}_{n_x} \\ -\mathbf{D}^T(\be)& -\mathbf{B}^T(\be) \\ \mathbf{I}_{n_y} & \mathbf{0}_{n_y \times n_x} \end{bmatrix}, \\
\label{eq:Mhalfthing5593}
&\mathbf{M}^{1/2}(z) = \left(  \sum_{k=1}^{N}  \sum_{j=1}^{m}    \sum_{\ell=1}^{\numJSmooth{k}}w_{k:k+1|N}^\ell (\be,j)  \expected_{k}^{\be,j,\ell} \left[\begin{bmatrix}X_k\\X_{k+1}\\u_k\\Y_k\end{bmatrix}\begin{bmatrix}X_k\\X_{k+1}\\u_k\\Y_k\end{bmatrix}^T\right] \right)^{1/2},
\end{align}
\end{subequations}
\end{small}
where a QR decomposition will be required to be undertaken on $\boldsymbol{\Pi}^{{1/2}}(\be)$ for the UT form. 
The square-root factor of the sum in \eqref{eq:Mhalfthing5593} can be calculated in a numerically stable way, as $C = (w_a A^TA + w_b B^TB + \dots)^{1/2}$ can be computed by the Q-less QR decomposition 
\begin{align}C = \mQb \begin{bmatrix} \sqrt{w_a}A \\ \sqrt{w_b}B \\ \vdots \end{bmatrix}.\end{align}
The equations for computing $\boldsymbol{\Gamma}(z)$ and $\mathbf{T}$ are unchanged from Lemma~\ref{lem:JMLS_EM_soln345}, however Lemma~\ref{lem:expectationNormal} is not required to compute the expectation of sufficient statistics $\boldsymbol{\Sigma}(z)$, $\boldsymbol{\Phi}(z)$, and $\boldsymbol{\Psi}(z)$. 
Instead it is convenient to use the newly formed $\boldsymbol{M}^{1/2}(z)$ with transformations to arrange the required parts of the combined expectation.
The equations for completing this are
\begin{subequations}
\begin{align}
\boldsymbol{\Sigma}(z) &= T_1 \left(\mathbf{M}^{1/2}(z)\right)^T\mathbf{M}^{1/2}(z)T_1^T, \\
\boldsymbol{\Phi}(z) &= T_2 \left(\mathbf{M}^{1/2}(z)\right)^T\mathbf{M}^{1/2}(z)T_2^T, \\
\boldsymbol{\Psi}(z) &= T_2 \left(\mathbf{M}^{1/2}(z)\right)^T\mathbf{M}^{1/2}(z)T_1^T, \\
T_1 &= \begin{bmatrix}\mathbf{I}_{n_x} & \mathbf{0}_{n_x}& \mathbf{0}_{n_u}&\mathbf{0}_{n_y} \\
\mathbf{0}_{n_x} & \mathbf{0}_{n_x}& \mathbf{I}_{n_u}&\mathbf{0}_{n_y},
\end{bmatrix},\\
T_2 &= \begin{bmatrix}\mathbf{0}_{n_x} & \mathbf{0}_{n_x}& \mathbf{0}_{n_u}&\mathbf{I}_{n_y} \\
\mathbf{0}_{n_x} & \mathbf{I}_{n_x}& \mathbf{0}_{n_u}&\mathbf{0}_{n_y}
\end{bmatrix}.
\end{align}
\end{subequations}
\end{lem}
\end{small}

\subsection{Algorithm overview}
For clarity, an overview of the operation of the proposed algorithm is provided within Algorithm~\ref{alg:JMLSEMalg34433}.
\begin{algorithm}
\caption{The Numerically stable JMLS EM algorithm}
\label{alg:JMLSEMalg34433}
\begin{algorithmic}[1]
\Require The initial guess of JMLS system parameters $\{\boldsymbol{\Pi}(z),\boldsymbol{\Gamma}(z)\}_{z=1}^{m}$, defined model transition matrix $\mathbf{T}$, measurement vector $y_k$, and exogenous input vector $u_k$ for time steps $k=1,..,N$. The statistics for the initial guess of the prior $p(x_1,z_1)$ is also required. 
\For{i=1,\dots}
\State Calculate the statistics of the forward filtered distribution $p(x_k,z_k|y_{1:k})$ for $k=1,\dots,N$ with the instructions provided within \cite{generalJMLSpaperBalenzuela}, each time step will require the use of \eqref{eq:conversionsystem378} to convert between systems. 
\State Calculate the statistics of the backwards likelihood $p(y_{k:N}|x_k,z_k)$ for $k=1,\dots,N$ using the instructions within \cite{generalJMLSpaperBalenzuela}, each time step will require the use of \eqref{eq:conversionsystem378} to convert between system conventions.
\State Calculate the statistics of the joint-smoothed distribution $p(x_{k+1},z_{k+1},x_k,z_k|y_{1:N})$ for $k=1,\dots,N$ using Lemma~\ref{lem:jointsmoothedmain43}.
\State Compute the new parameter estimates using the equations provided in Lemma~\ref{sec:M-stepqrt346} and the supporting Lemma~\ref{lem:sqrtmassExpect45768}.
\EndFor
%
\end{algorithmic}
\end{algorithm}
\clearpage
\section{Discussion}
\label{sec:EMconsider4}
In this section we discuss drawbacks, and additional considerations with using the EM algorithm for identification of JMLS systems.
Many of the issues discussed here are general properties of the EM algorithm, and are not limited to the proposed method.
\subsection{Stopping criterion}
A good indicator of convergence to a solution is noting when the log-likelihood 
\begin{align}
\label{eq:lnLJMLSsystem453}
&\ln\ptp(y_{1:N}) \nonumber \\
&= \sum_{k=1}^{N} \ln \left( \sum_{z_k=1}^m\int \ptp(y_k|x_k,z_k) \ptp(x_k,z_k|y_{0:k-1}) \, dx_k \right) \nonumber \\
&= \sum_{k=1}^N \ln \left( \sum_{z_k =1}^{m}\sum_{i=1}^{\numPred{k}} \tilde{w}^i_{k | k}(z_k) \right),
\end{align}
does not exceed a threshold for a number of EM iterations, where $\tilde{w}^i_{k | k}(z_k)$ is the un-normalised weights within operation of the JMLS filter, and $M^\text{p}_k$ is the number of components in the predicted mixture at time step $k$.
Note that this is often the approximated log-likelihood, as the true log-likelihood has an exponential cost to generate and is computationally intractable to do so for more than a few time steps \cite{blackmore2007model}.
In practice, log-weights should be used, and the log-likelihood calculation can make use of the log-sum-exponent trick.
%
\subsection{Equivalent systems}
It is well known that for a linear system, the state-space (SS) realisation is not unique, and there are an infinite number of state-space systems which have the same input-output behavior \cite{bako2009identification,vidal2002observability,svensson2014identification}.
This creates difficulty when validating an EM algorithm; for a linear system some options are to compare the transfer functions, which can be generated from a SS system, or compare the Bode plot from both systems.

For a JMLS system the model parameters $\theta(z_k)$ can be transformed into $\bar{\theta}(z_k)$ using the following formulas with no change to input output behavior,
\begin{align}
\bar{\mathbf{A}}(z) = \mytrans\mathbf{A}(z)\mytrans^{-1},
\bar{\mathbf{B}}(z) = \mytrans\mathbf{B}(z), \nonumber \\
\bar{\mathbf{C}}(z) = \mathbf{C}(z)\mytrans^{-1}, 
\bar{\mathbf{D}}(z) = \mathbf{D}(z), \nonumber \\
\bar{\mathbf{Q}}(z) = \mytrans\mathbf{Q}(z)\mytrans^T, 
\bar{\mathbf{S}}(z) = \mytrans\mathbf{S}(z),
\bar{\mathbf{R}}(z) = \mathbf{R}(z).
\end{align}
Where $\mytrans$ is any real invertible matrix of the required dimension ($n_x\times n_x$).
In general, if a minimal number of models is used, the same linear state transformation $\mytrans$ must be applied to each of the models $z=1,\dots,m$.

This problem is made more complex when comparing JMLS systems, as the models may not be found in the same order as the true system, i.e., model 1 and 2 may be swapped.
The order in which models are found is \emph{not} important, and the system model index and parameters can be remapped to yield an equivalent system \cite{svensson2014identification}, i.e. $\mathbf{A}(z) = \bar{\mathbf{A}}(M(z))$, where $M(z) \neq z$.
For this reason it is only necessary to consider the transition probabilities between such models.
One way to identify the mapping between the identified and ground truth models $M(\cdot)$ is to consider the $l^2$ error from the magnitude Bode plot with each of the possible 1-to-1 mappings.
\subsection{Convergence}
While the EM algorithm is guaranteed to converge to local minimum \cite{balakrishnan2004inference}, it is well known that the EM approach is in general not globally convergent \cite{logothetis1999expectation}.
With application to estimation of jump Markov linear systems, the algorithm can converge to a local minimum where some models are disabled by zeros in elements of the state transition matrix $\mathbf{T}$.
This is a problem encountered in literature previously.

Some available approaches to this problem include a numerical hack, where zeros in the transition matrix are replaced  with a small value $\epsilon$ to keep these hypothesizes alive \cite{jilkov2004online}.
Alternatively restart methods can be employed, where the algorithm is re-run using a new initial parameter set, which is sampled according to a distribution, or from using GM clustering to indicate unexplored regions \cite{gil2009beyond}. However, neither of these approaches are practical if the EM algorithm takes too long to converge.

One method which we have practiced to improve convergence is to ensure the linear model parameters have converged, before estimating the transition matrix.
This avoids unlikely models being turned off forever.
Additionally, \textit{do not} start the algorithm with identical models, as this will cause the algorithm to become stuck in a local maxima, as per the reasoning with the following lemma.

\begin{lem}
\label{lem:samemodelparaminit}
Initialising multiple modes with the same parameter set, prior, and equal transition probabilities $\textbf{T}(z_{k+1},z_k)=\frac{1}{m}$ will resort in a common update to the parameters in the next iteration of the EM algorithm.
This will cause the parameter sets to be identical for each subsequent iteration of the algorithm, and effectively produce a system with an incorrect number of models.
\end{lem}

\subsection{Observability}
It is important for the identifiability of the system, that the series of data not only excite each of the linear models, but also excite the switching condition \cite{jilkov2004online}.
As obervability of switched systems has been covered thoroughly in the literature, we do not discuss it further, and instead direct the interested reader to \cite{bako2009identification,vidal2002observability}.

\clearpage
\section{Simulations}
\label{sec:simulations5}
Here we provide the simulation results from identifying various jump Markov linear systems to demonstrate the effectiveness of the proposed solution.
\subsection{Identification of a single state jump Markov linear system}
In this example we consider a JMLS system used in \cite{helmick1995fixed,kim1994dynamic,doucet2001particle,barber2006expectation,svensson2014identification} with the form
\begin{subequations}
\label{eq:differentJMLSconventions4}
\begin{align}
&X_{k} = \mathbf{A}(Z_k)X_{k-1} + \mathbf{B}(Z_k)u_{k} + V_{k-1}(Z_k), \\
&Y_{k} = \mathbf{C}(Z_k)X_k + \mathbf{D}(Z_k)u_k + E_k(Z_k), \\
&V_{k-1}(Z_k) \sim \mathcal{N}({0},\mathbf{Q}(Z_k)), \\
&E_k(Z_k) \sim \mathcal{N}({0},\mathbf{R}(Z_k)). 
\end{align}
\end{subequations}
Where the proposed algorithm requires a minor modification for this model convention, as cross-covariance $\mathbf{S}(z_k)$ is not allowed, and a different time index is attached to the discrete random variable and input in the process model.

This simple single state system was chosen as higher state dimensions require an increased particle count within the PSEM and PSAEM algorithms, and hence would require an unnecessary large computational time.
The system used for this example was parameterised by
\begin{align}
\mathbf{A}(1)&=0.9, \mathbf{B}(1)=0.1, \mathbf{C}(1)=0.9, \mathbf{D}(1)=0, \nonumber \\
\mathbf{Q}(1) &= 0.045, \mathbf{R}(1) = 0.002, \nonumber \\
 \mathbf{A}(2)&=0.65, \mathbf{B}(2)=-0.32, \mathbf{C}(2)=1, \mathbf{D}(2)=0, \nonumber \\
\mathbf{Q}(2) &= 0.002, \mathbf{R}(2) = 0.005, \nonumber \\
\mathbf{A}(3)&=0.51, \mathbf{B}(3)=0.2, \mathbf{C}(3)=1.2, \mathbf{D}(3)=0, \nonumber \\
\mathbf{Q}(3) &= 0.02, \mathbf{R}(3) = 0.009, \nonumber \\
\mathbf{T} &= \begin{bmatrix} 0.6 & 0.35 & 0.1 \\ 0.3 & 0.6 & 0.4 \\ 0.1 & 0.05 & 0.5 \end{bmatrix}.
\end{align}
The system was simulated for $N=7000$ time steps, with $u_k \sim \mathcal{N}(0,1)$, and $x_0 = 0$. The measurements and inputs of this series were then issued to the EM algorithms for comparison of performance.
The initial guess given to the EM algorithms were
\begin{align}
\hat{\mathbf{A}} =1.2 \mathbf{A}, \hat{\mathbf{B}} = 0.8 \mathbf{B}, \hat{\mathbf{C}} = 0.5  \mathbf{C}, \hat{\mathbf{D}} = 0, \nonumber \\
\hat{\mathbf{Q}} = 0.8  \mathbf{Q}, \hat{\mathbf{R}}=1.5 \mathbf{R}, \nonumber \\
\hat{\mathbf{T}} = \frac{1}{3} \mathbf{1}_3,
\end{align}
where $\mathbf{1}_n$ denotes a $n\times n$ ones matrix, and a hat on the parameter denotes an initial guess.
Additionally, the prior used for each method was $p(x_0,z_0)=\frac{1}{3}\mathcal{N}(x_0|0,1)$.

The EM algorithms compared in this example were 
\begin{itemize}
\item The nonlinear particle PSEM \cite{schon2011system} with 75 particles and 34 trajectories.
\item Stochasic approximation PSAEM \cite{lindsten2013efficient} with a particle count of 118.
\item Rao-Blackwellized stochastic approximation method RB-PSAEM \cite{svensson2014identification} with 40 trajectories.
\item The proposed method JMLS-EM using KLR reduction to approximate distributions as a weighted GM with 3 components per discrete mode.
\end{itemize}
As each of the methods use different approximations, with different approximation settings, the methods can only be compared fairly on computational time.
Because of this, the above algorithms were each allowed 500 EM iterations, with the above settings chosen such that the EM algorithms took approximately 55 seconds to complete an iteration.

\begin{figure}
\begin{center}
\includegraphics[width=0.8\columnwidth]{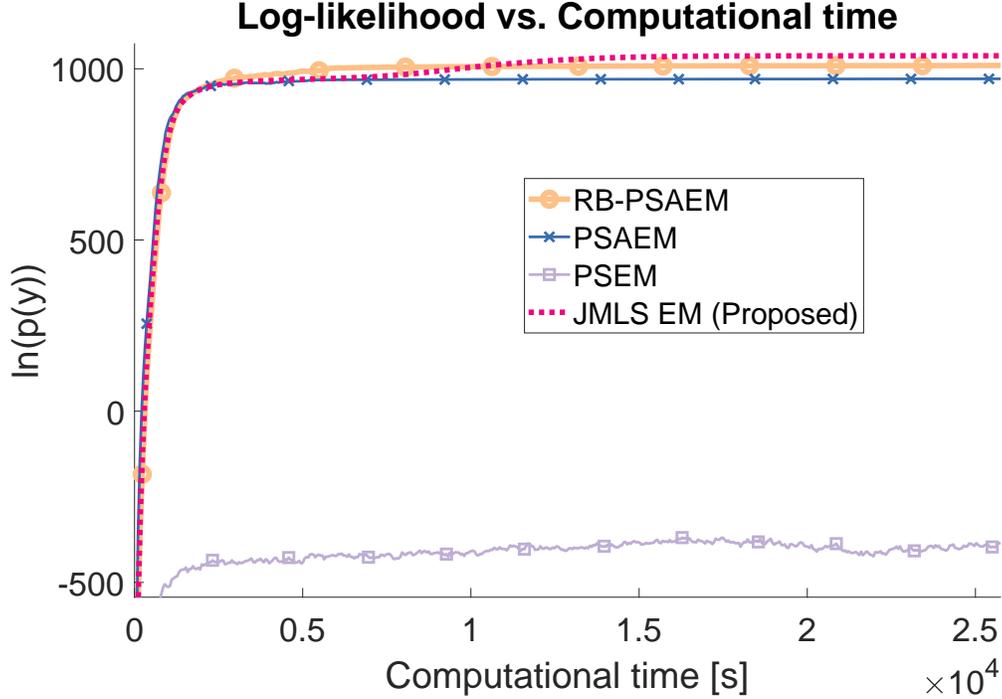}    
\caption{Approximated log-likelihood of the estimated parameter set at each iteration of the EM algorithms. The alternate RB-PSAEM method (solid yellow with circles), PSAEM method (solid blue with crosses), PSEM method (solid light purple with squares), can be compared to the proposed method, shown in dotted magenta.}  
\label{fig:exp_one_lnL}                                 
\end{center}                                 
\end{figure}

Figure~\ref{fig:exp_one_lnL} shows the approximate log-likelihood of the parameter estimate at each iteration for each of the methods.
The log-likelihood was computed by running a JMLS filter with a KLR merging strategy, where each discrete mode within the filter was allowed to store a weighted Gaussian mixture with 5 components.
This increase in components was used to remove some of the bias toward the proposed method, as it uses the same merging approximation, and was not able to be increased further due to computational time.
%

\begin{figure}
\begin{center}  
     \subfloat[Frequency response from the $1_\text{st}$ linear model\label{subfig-1:dummy}]{%
       \includegraphics[width=0.45\textwidth]{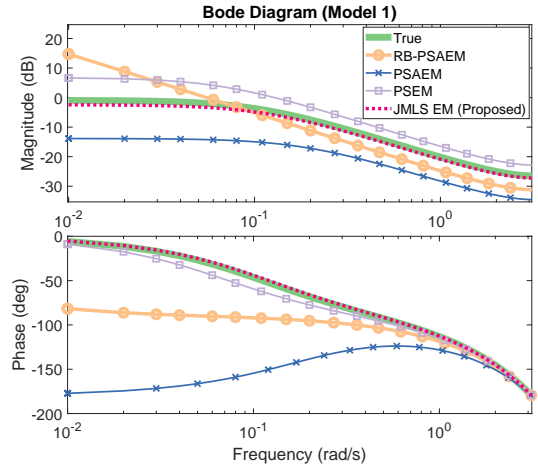}
     }
     \vfill
     \subfloat[Frequency response from the $2_\text{nd}$ linear model\label{subfig-2:dummy}]{%
       \includegraphics[width=0.45\textwidth]{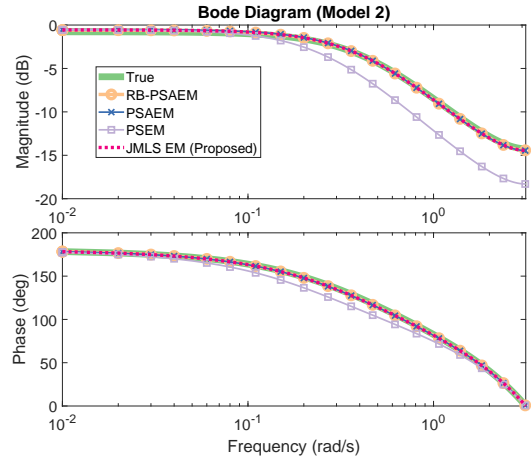}
     }
     \vfill
     \subfloat[Frequency response from the $3_\text{rd}$ linear model\label{subfig-2:dummy1}]{%
       \includegraphics[width=0.45\textwidth]{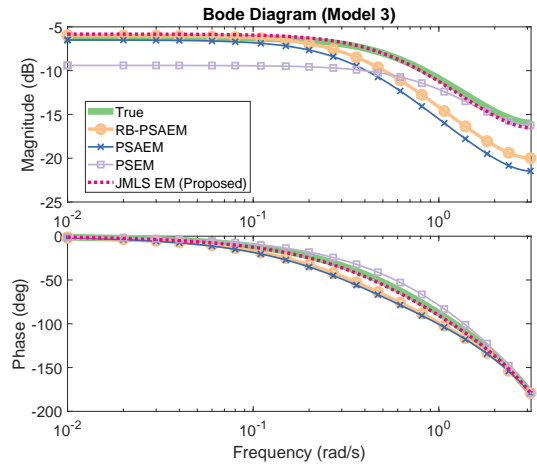}
     }
     \caption{Frequency response from the three first order models, where the truth (solid green) is shown along with the competing methods, and the proposed method (dotted magenta).}
\label{fig:exp_one_bode}
\end{center}  
\end{figure}

The frequency response of each of the modes identified from each method is shown by the Bode plots within Figure~\ref{fig:exp_one_bode}.

\subsection{Identification of a second order jump Markov linear system}
\label{sec:second_expspecs}
In this example we consider the case of identifying a system with a higher state dimension. 
As the PSAEM and PSEM methods utilise $1_\text{st}$ order particle approximations, which are well known to suffer from the \textit{curse of dimensionality}, they have been discounted from further examples.
Hence, this example will consider the RB-PSAEM and proposed method only.
As with Example 1, the proposed algorithm required a minor modification.

The target system can be described by
\begin{subequations}
\label{eq:system3534453}
\begin{align}
\mathbf{T} &= \begin{bmatrix} 0.6 & 0.5 \\ 0.4 & 0.5\end{bmatrix}, \\
H_1(\text{z}) &=\frac{0.7406 \text{z} + 0.004861}{\text{z}^2 + 0.6178 \text{z} + 0.4385}, \\
H_2(\text{z}) &= \frac{-1.461 \text{z} + 1.98}{\text{z}^2 - 1.189 \text{z} + 0.2715},
\end{align}
\end{subequations}
which was randomly generated. 
Note that the $\text{z}$ used within \eqref{eq:system3534453} is the variable from application of the z-transform to a transfer function, and differs from the $z$ used previously, which indicated the mode of operation of the system.
This system was simulated for $N=7000$ time steps, with $u_k \sim \mathcal{N}(0,1)$, and $x_0 = \vec{0}$.

\begin{figure}
\begin{center}
\includegraphics[width=0.8\columnwidth]{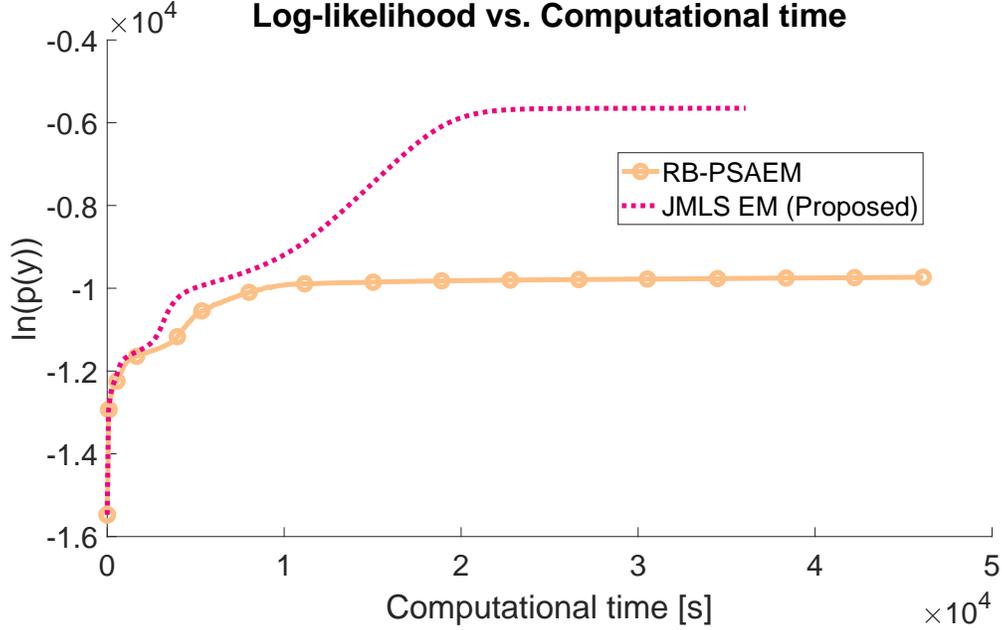}    
\caption{Approximated log-likelihood of the estimated parameter set at each iteration of the EM algorithms. The proposed method is shown in dotted magenta, where the alternate RB-PSAEM method is shown in solid yellow with circles.}  
\label{fig:ex2_likelihood}                                 
\end{center}                                 
\end{figure}

The prior used for both methods was $p(x_0,z_0)=\frac{1}{2}\mathcal{N}(x_0|\vec{0},\mathbf{I}_2)$, with an initial system guess of another randomly generated system and $\hat{\mathbf{T}}=\frac{1}{2}\mathbf{1}_2$.
The RB-PSAEM algorithm was allowed 25 trajectories and the proposed algorithm was allowed 3 components per discrete state.
These algorithms each ran for 500 EM iterations, with the RB-PSAEM method taking longer than the proposed method.
The approximated log-likelihood of the estimated set of parameters per iteration is shown in Figure~\ref{fig:ex2_likelihood} for both approaches. 
However, it should be noted that the RB-PSAEM method is still slowly converging.

\begin{figure}
\begin{center}
     \subfloat[Frequency response from the $1_\text{st}$ linear model\label{subfig-1:ex2a}]{%
       \includegraphics[width=0.65\textwidth]{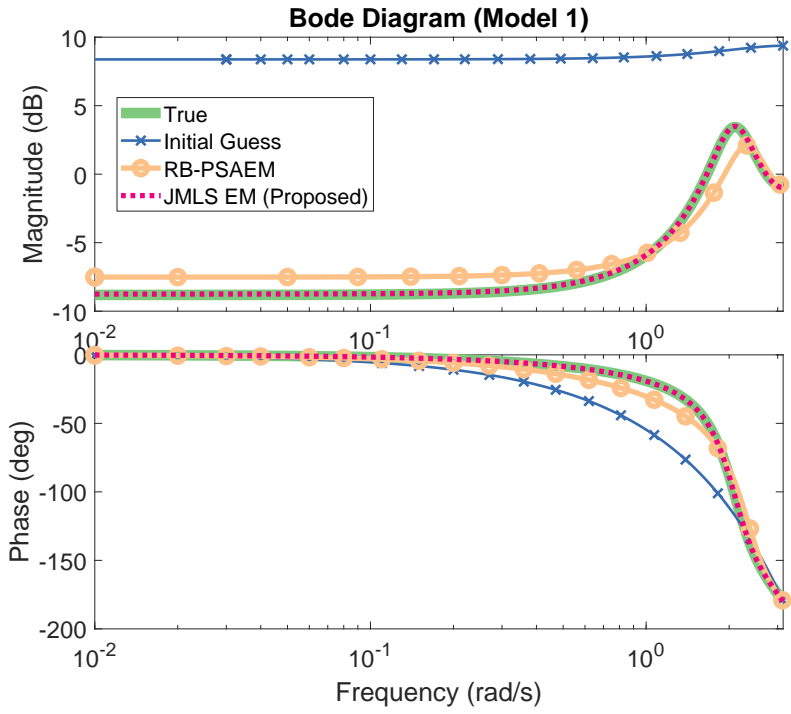}
     }
     \vfill
     \subfloat[Frequency response from the $2_\text{nd}$ linear model\label{subfig-2:ex2b}]{%
       \includegraphics[width=0.65\textwidth]{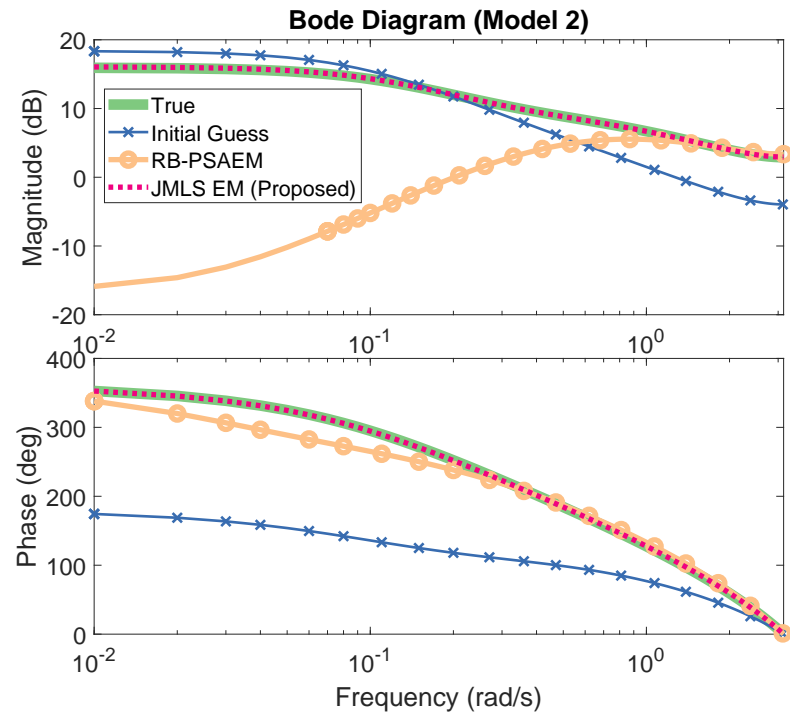}
     }
     \caption{Frequency response from the two second order models, where the truth (solid green) is shown along with the initial guess (solid blue with crosses), competing RB-PSAEM method (solid yellow with circles), and the proposed method (dotted magenta)}
\label{fig:exp_twoa_bode}
\end{center}
\end{figure}

The frequency response of the identified system, true system and initial system guess are shown within the Bode plots in Figure~\ref{fig:exp_twoa_bode}.

\subsection{Dynamic JMLS identification}

In this example we consider the dynamic JMLS system operating according to \eqref{eq:JMLSdef1}. 
As alternative methods do not use this convention, and instead operate according to \eqref{eq:differentJMLSconventions4}, this example cannot compare to them.

To demonstrate the practicality of the proposed algorithm, a $5_\text{th}$ order single input single output (SISO) system with two modes was randomly generated, with the transfer functions
\begin{small}
\begin{subequations}
\begin{align}
&H_1(\text{z})= \nonumber \\ &\frac{ -2.056 \text{z}^4 + 1.232 \text{z}^3 + 0.01248 \text{z}^2 + 0.2513 \text{z} - 0.142}{\text{z}^5 - 1.173 \text{z}^4 + 0.3133 \text{z}^3 - 0.07036 \text{z}^2 + 0.1083 \text{z} - 0.03329},\\
&H_2(\text{z}) =\nonumber \\ &\frac{1.749 \text{z}^4 - 1.185 \text{z}^3 - 0.2067 \text{z}^2 + 0.0918 \text{z} + 0.01656}{\text{z}^5 - 0.224 \text{z}^4 - 0.1758 \text{z}^3 - 0.007425 \text{z}^2 + 0.01095 \text{z} + 0.001546},
\end{align}
\end{subequations}
\end{small}

which was simulated for $N=2000$ time steps, with \mbox{$x_0 = \vec{0}$}.

Afterwards, the proposed EM method was ran on the dataset, initialised by another set of random $5_\text{th}$ order models, a prior of $p(x_0,z_0)=\frac{1}{2}\mathcal{N}(x_0|\vec{0},3\mathbf{I}_5)$, and the initial state transition matrix $\hat{\mathbf{T}}=\frac{1}{2}\mathbf{1}_2$.
The KL reduction stage within the proposed algorithm was allowed to keep six components per discrete state.
Estimation of the transition probabilities was enabled when the log-likelihood change was less than 0.03 for 10 consecutive time steps.
Finally, convergence to a solution took 1418 iterations of the proposed EM algorithm.

\begin{figure}
\begin{center}
\includegraphics[width=\columnwidth]{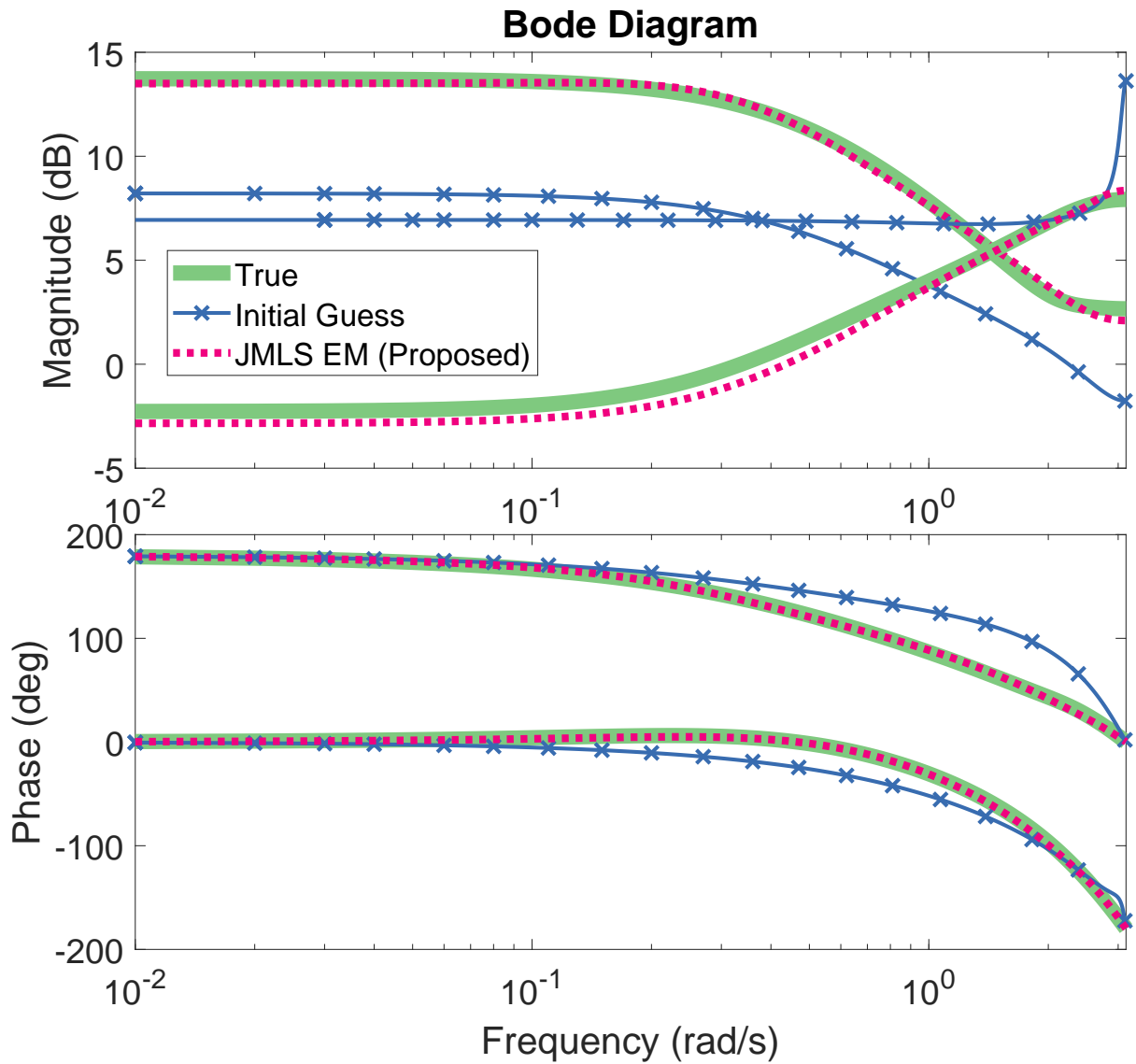}    
\caption{Frequency response from the two $5_\text{th}$ order dynamic models. The true system is shown in solid green, where the blue trend with crosses marks the initial guess, and finally the identified system from the proposed method is shown in dotted magenta. }  
\label{fig:fith_order_dyn}                                 
\end{center}                                 
\end{figure}

The frequency response from the identified modes, initial guess and true system are shown within Figure~\ref{fig:fith_order_dyn}.
The true and identified transition matrix was
%
\begin{align}
\mathbf{T}_\text{true} = \begin{bmatrix}0.7 & 0.35\\ 0.3 & 0.65 \end{bmatrix}, \text{ and } \mathbf{T}_\text{identified} = \begin{bmatrix}0.6922 &   0.3814 \\
    0.3078   &  0.6186 \end{bmatrix},
    \end{align}
respectively.

\subsection{Demonstration of local convergence}
In this example, we consider the case where the proposed algorithms were initialised with a poor choice of parameters, and converged to a local maxima solution.

The system used in this example operating according to the JMLS convention outlined in Example 1, requiring the modification to the proposed algorithm, and was parameterised by
\begin{subequations}
\begin{align}
\mathbf{T}&= \begin{bmatrix} 0.6 & 0.05& 0.4\\0.3& 0.6& 0.1 \\ 0.1& 0.35& 0.5\end{bmatrix},\\
H_1(\text{z}) &= \frac{-0.3458 \text{z}^2 - 0.5586 \text{z} - 0.0779}{\text{z}^3 + 0.7556 \text{z}^2 + 0.0832 \text{z} + 0.001395},\\
H_2(\text{z}) &= \frac{0.06533 \text{z}^2 - 0.2209 \text{z} - 0.04806}{\text{z}^3 - 1.087 \text{z}^2 + 0.5054 \text{z} - 0.1482},\\
H_3(\text{z}) &= \frac{-0.06093 \text{z}^2 - 0.03161 \text{z} + 0.007855}{\text{z}^3 + 0.1836 \text{z}^2 + 0.003266\text{z} - 0.00178},
\end{align}
\end{subequations}

\begin{figure}
\begin{center}
\includegraphics[width=0.5\columnwidth]{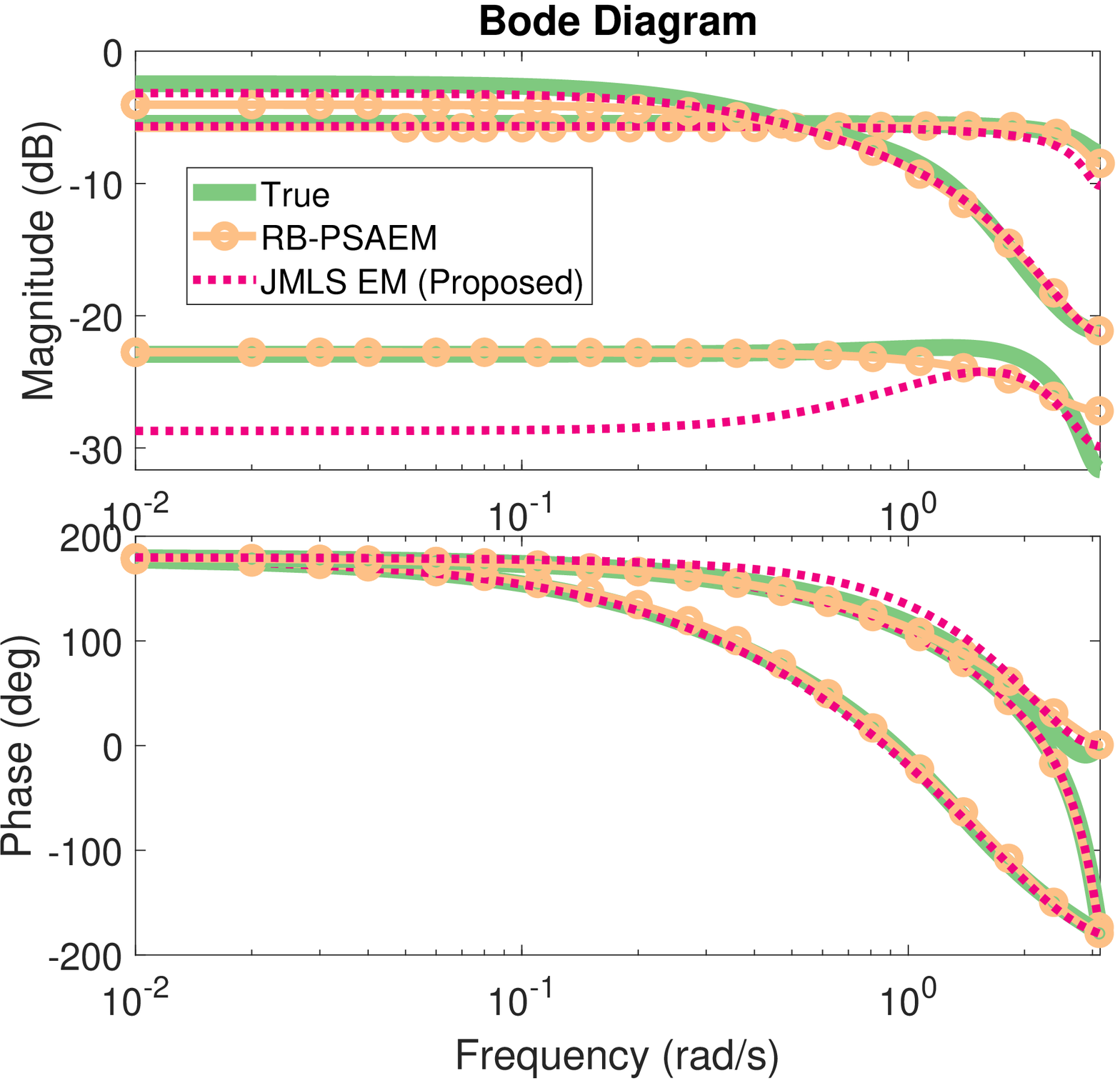}    
\caption{Frequency response of the converged solution with a poor initial guess. Where the true system has the response shown in solid green, the RB-PSAEM method provided a system with the response shown in solid yellow with circles, and the proposed method provided a system with the response shown in dotted magenta.}  
\label{fig:poorguyess33}                                 
\end{center}                                 
\end{figure}

\begin{figure}
\begin{center}
\includegraphics[width=0.5\columnwidth]{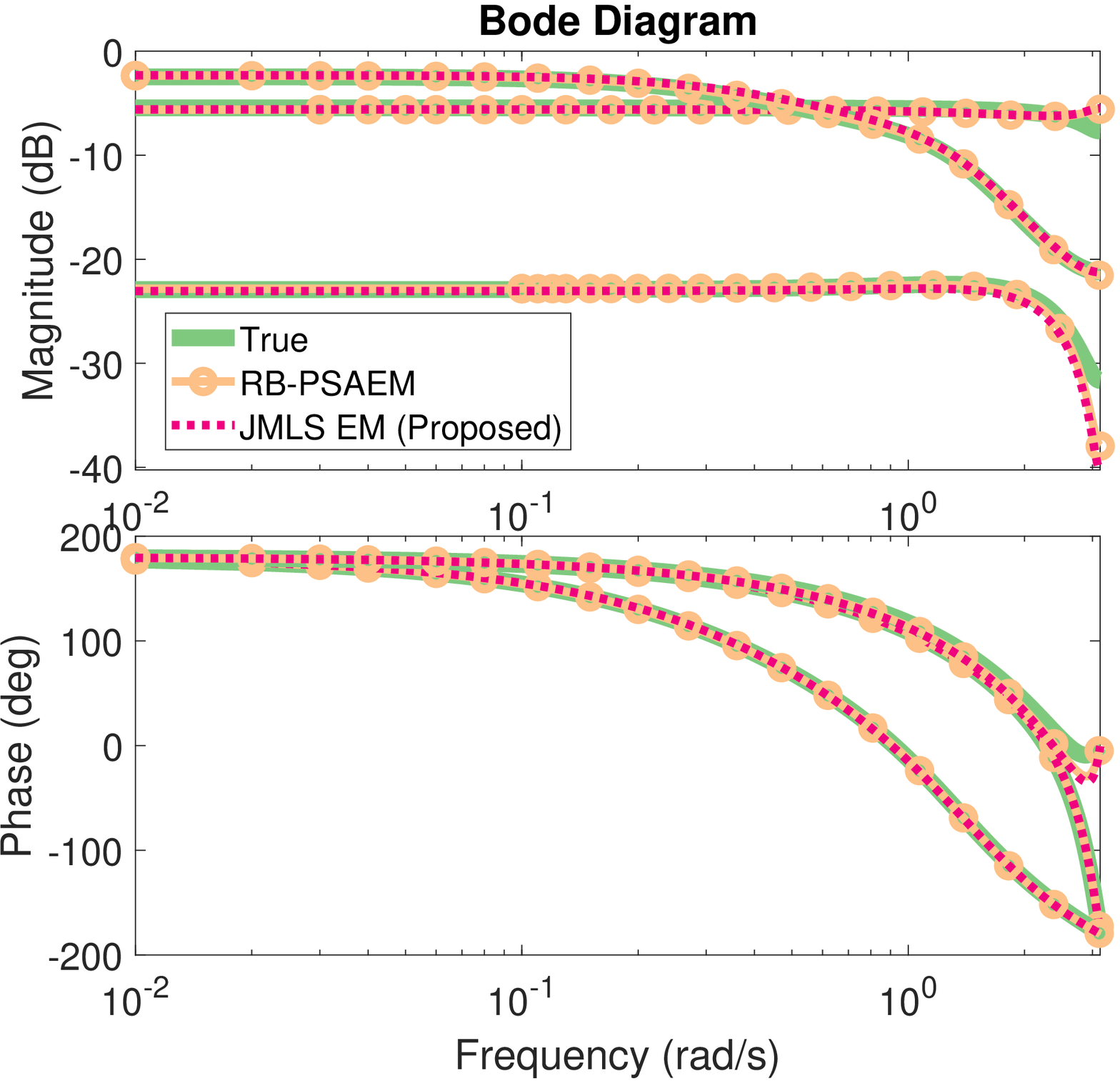}    
\caption{Frequency response of the converged solution with an initial guess of the true parameters. Where the true system has the response shown in solid green, the RB-PSAEM method provided a system with the response shown in solid yellow with circles, and the proposed method provided a system with the response shown in dotted magenta.}  
\label{fig:goodguess34}                                 
\end{center}                                 
\end{figure}

The system was simulated for $N=8000$ time steps with the system input $u_k \sim \mathcal{N}(0,1)$, and the initial condition $x_0 = \vec{0}$.
Both the proposed algorithm and the RB-PSAEM algorithm were allowed 300 EM iterations to converge to a solution, with each iteration taking approximately 835 seconds.
The proposed algorithm was allowed 4 components per discrete state, whereas the RB-PSAEM method was allowed 120 trajectories.
Both of these methods were initialised with the prior
\begin{align}
p(x_0,z_0) = \frac{1}{3}\mathcal{N}(x_0|\vec{0},\mathbf{I}_3).
\end{align}
The frequency response of the converged system with the poor initial guess is shown within Figure~\ref{fig:poorguyess33}, whereas by starting from the true values, the algorithms converge to values providing the frequency response shown in Figure~\ref{fig:goodguess34}. 

In this example the randomness within the RB-PSAEM algorithm appears to have aided its' convergence close to the true solution, where the proposed method has identified a local maximum, which is a well known problem with the EM algorithm.
As shown by Figure~\ref{fig:goodguess34}, the proposed algorithm is capable of converging to the global maximum, if initialised with a good guess of parameter values. 
%
%
%
%
\subsection{Robustness testing}
\begin{figure}
\begin{center}  
     \subfloat[Frequency responses from the identified models using the RB-PSAEM method.]{%
       \includegraphics[width=0.65\textwidth]{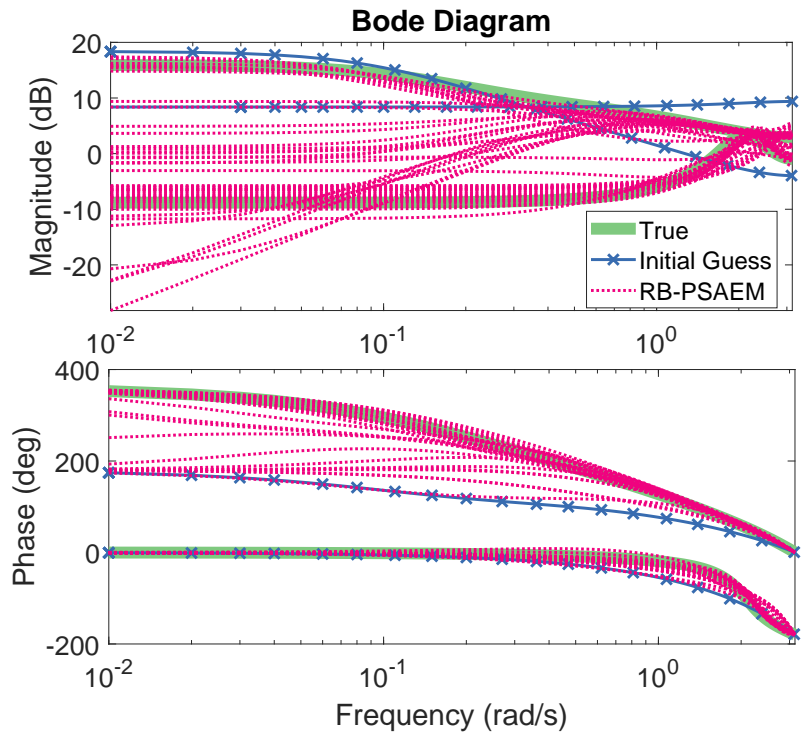}
     }
     \vfill
     \subfloat[Frequency responses from the identified models using the proposed method.]{%
       \includegraphics[width=0.65\textwidth]{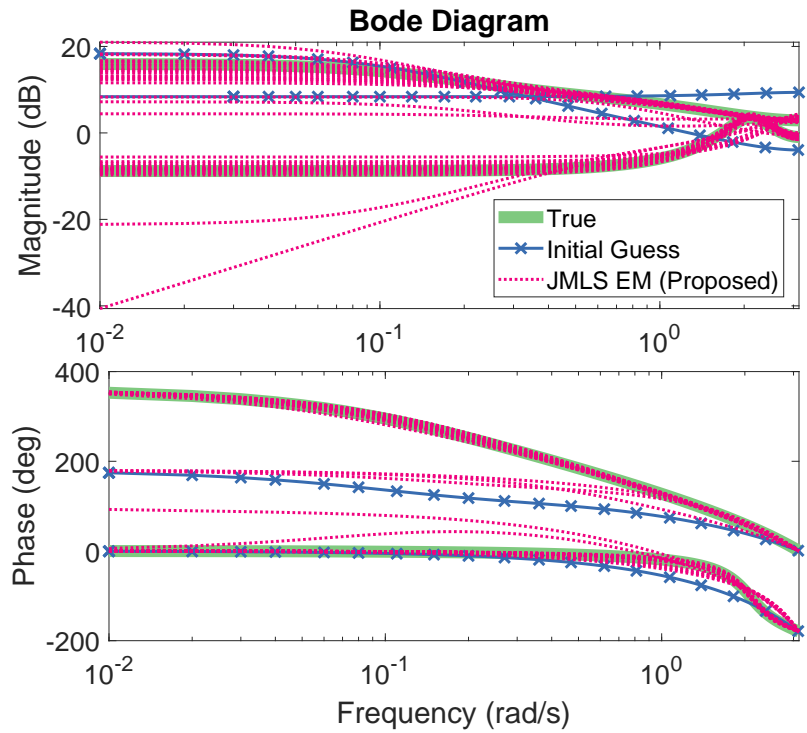}
     }
     \caption{Frequency response of the models, where the truth (solid green) is shown along with the initial guess (solid blue with crosses), and the identified models (dotted magenta).}
\label{fig:exp_5_bode_elaire_fortune}
\end{center}  
\end{figure}

In this example, the two-mode two-state system from subsection~\ref{sec:second_expspecs} was used to generate 25 different datasets, each with a length of $N=250$ time steps.
On each of these datasets, the proposed method and the alternate RB-PSAEM method was allowed 500 EM iterations, with both methods taking about 1400 seconds to complete.
Figure~\ref{fig:exp_5_bode_elaire_fortune} shows the frequency response response of the identified systems from each run, the initial guess common to each run, and the true system.

In the majority of runs, the proposed method seems to have accurately captured the system dynamics, whereas the RB-PSAEM method had not yet converged for many of these runs.   
In the few cases, where the proposed method performed poorly, the RB-PSAEM method also appears to have suffered, which is believed to be due to the short dataset.
\clearpage
\section{Conclusion}\label{sec:conclusion}
From the examples given, we have demonstrated that the proposed variational method is a practical algorithm, which can outperform its' competitors when given a reasonable computational time. 
This however, relies on the merging procedure used within proposed method being allowed to store enough components to describe the smoothed distribution with sufficient accuracy. Which, in turn, allows the M-step to compute parameters close to what an exact EM implementation would.

JMLS system identification is a computationally expensive problem, scaling exponentially with the number of time steps.
We have presented another alternative approximation to this problem, which is an SMC-free algorithm.
As merging based methods do not have statistical convergence properties, using too few of modes will cause the algorithms to arrive at a sub-optimal solution. 
Increasing the number of components allowed in the joint smoothed distribution, and checking convergence to a similar set of values would be one way to ensure this approximation has not affected the outcome.

Admittedly, the alternative methods using the stochastic approximation have convergence guarantees. However, as shown from the examples, these can be very slow to converge, and take an impractical amount of time. 
This is most prominent in Example 2, where the log-likelihood is still increasing at a rate that looks almost flat in Figure~\ref{fig:ex2_likelihood}.

Finally, we conclude that without using the exact $\mathcal{Q}$-function, generated from an exact smoothed distribution, an increase in the log-likelihood for each EM iteration cannot be guaranteed by the proposed method nor by any available competing method.
%
%
%
%
%
%
%
\section{Acknowledgements}
We would like to thank Dr. Andreas Svensson for kindly providing MATLAB code for the alternate methods.

\clearpage
\appendix
\section{Joint smoother Lemmata}
In the following Lemmata we use the shorthand
\begin{align}
\chi_k= \begin{bmatrix}x_k\\x_{k+1}\end{bmatrix}
\end{align}
\begin{small}
\begin{lem}
The square-root
$\mathbf{P}^{1/2}_{k:k+1|N} = \mR_{22}$
where $ \mR_{22}$ is a component from the Q-less QR decomposition of
\begin{align}\begin{bmatrix} \mR_{11} & \mR_{12} \\0 & \mR_{22} \end{bmatrix}=\mQb \begin{bmatrix} \mathbf{I} & \mathbf{0} \\ \mathbf{J}_k & \mathbf{P}_{k:k+1|k}^{1/2}  \end{bmatrix},\end{align}
and
$$\mathbf{J}_k = \mathbf{P}_{k:k+1|k}^{1/2} (\mathbf{L}_{k:k+1|k+1}^{1/2})^T.$$
Note that the left half of $\mathbf{P}_{k:k+1|k}^{1/2} (\mathbf{L}_{k:k+1|k+1}^{1/2})^T$ will be the zeros matrix.
\end{lem}
\begin{pf}
From Lemma~\ref{lem:jointsmoothedmain5}, we know the joint-covariance is constructed using the line below, then applying the Woodbury matrix identity yields the following
\begin{align} &\mathbf{P}_{k:k+1|N} =\big(\mathbf{P}_{k:k+1|k}^{-1} + \mathbf{L}_{k:k+1|k+1} \big)^{-1} \nonumber \\
&= \mathbf{P}_{k:k+1|k}  - \mathbf{P}_{k:k+1|k} (\mathbf{L}_{k:k+1|k+1}^{1/2})^T  \nonumber \\
&\cdot (\mathbf{I} + (\mathbf{L}_{k:k+1|k+1} )^{1/2} \mathbf{P}_{k:k+1|k} (\mathbf{L}_{k:k+1|k+1}^{1/2} )^T)^{-1}  \nonumber \\
&\cdot (\mathbf{L}_{k:k+1|k+1}^{1/2})\mathbf{P}_{k:k+1|k}. \end{align}

Now consider the proposed Q-less QR decomposition of
\begin{align}\mRb=\mQb \begin{bmatrix} \mathbf{I} & \mathbf{0} \\ \mathbf{P}_{k:k+1|k}^{1/2} (\mathbf{L}_{k:k+1|k+1}^{1/2})^T & \mathbf{P}_{k:k+1|k}^{1/2}  \end{bmatrix},\end{align}
which has the following form and solutions,
$$\mRb  = \begin{bmatrix} \mR_{11} & \mR_{12} \\0 & \mR_{22} \end{bmatrix} = \mQb \begin{bmatrix} E &0\\F&G\end{bmatrix}\label{eq:qrform2},$$

$$\mR_{11}^T\mR_{11} =E^TE+F^TF = \mathbf{I} + (\mathbf{L}_{k:k+1|k+1} )^{1/2} \mathbf{P}_{k:k+1|k} (\mathbf{L}_{k:k+1|k+1}^{1/2} )^T,$$

\begin{align}
&\mR_{22}^T\mR_{22} 
= \mathbf{P}_{k:k+1|k} - \mathbf{P}_{k:k+1|k} (\mathbf{L}_{k:k+1|k+1}^{1/2})^T (\mR_{11}^T\mR_{11})^{-1} \nonumber \\
& \quad \cdot (\mathbf{L}_{k:k+1|k+1}^{1/2})\mathbf{P}_{k:k+1|k} 
=\mathbf{P}_{k:k+1|N}.
\end{align}
\end{pf}

\begin{lem}
\label{lem:jointsmoothedmain5}
Given the statistics of the forwards filter and and backwards filter in the form of
  \begin{align}
    p(x_k,z_k \mid y_{1:k}) &= \sum_{i=1}^{\numFilt{k}} w^i_{k | k}(z_k) \, \mathcal{N}  (x_k | \mu^i_{k
                              \mid k}(z_k), \, \myP^i_{k \mid k}(z_k)
                               ),
  \end{align}

  and
    \begin{align}
    \label{eq:MCHA6100 Notes on JMLS:292323}
    p&(y_{k:N} \mid x_{k},z_{k}) = \sum_{i=1}^{\numCorr{k}} \mathcal{L}  ( x_k  | \bar{r}_{k}^i(z_{k}),\, \bar{s}_{k}^i(z_{k}),\, \bar{\myL}_{k}^i(z_{k})  ),
\end{align}
 respectively, the joint smoothed distribution can be calculated as
  \begin{align}
  p&(x_{k+1},z_{k+1},x_k,z_k \mid y_{1:N}) =  \sum_{j=1}^{\numJSmooth{k}} w^j_{k:k+1 | N}(z_k,z_{k+1}) \nonumber \\
  & \cdot \mathcal{N}  ( \chi_k\, | \, \mu^j_{k:k+1
                              \mid N}(z_k,z_{k+1}), \, \myP^j_{k:k+1 \mid N}(z_k,z_{k+1})
                               ),
  \end{align}
 where
\begin{subequations}
\begin{align}
&\numJSmooth{k}  = \numFilt{k}\cdot \numCorr{k+1},\\
j &= \numFilt{k} (\ell - 1) + i,\\
{w}&_{k:k+1 | N}^j(z_k,z_{k+1})  \\
&= \frac{\tilde{w}_{k:k+1 \mid N}^j(z_k,z_{k+1})}{\sum_{z_{k+1}=1}^{m}\sum_{z_k=1}^{m} \sum_{p=1}^{\numJSmooth{k}} \tilde{w}_{k:k+1 \mid N}^p(z_k,z_{k+1})},\\
\tilde{w}&_{k:k+1 \mid N}^j(z_k,z_{k+1}) = e^{\frac{1}{2}\beta^j(z_k,z_{k+1})} ,\\ 
        \beta&^j(z_k,z_{k+1}) \nonumber \\
        &= (\mu^j_{k:k+1|N}(z_k,z_{k+1}))^T(\myP_{k:k+1|N}^j(z_k,z_{k+1}))^{-1}\mu^j_{k:k+1|N}(z_k,z_{k+1}) \nonumber \\
    &\quad -  (\mu^i_{k:k+1 \mid k}(z_k))^T (\myP^i_{k:k+1 | k}(z_k))^{-1} \mu^i_{k:k+1 | k}(z_k)  \nonumber \\
    &\quad  -\bar{r}^\ell_{k+1}(z_{k+1}) +\ln|\myP_{k:k+1|N}^j(z_k,z_{k+1})| - \ln|\myP_{k:k+1|k}^i(z_k)|  \nonumber \\
    &\quad + 2\ln(w^i_{k|k}(z_k)) + 2\ln(\mathbf{T}(z_{k+1},z_k)), \\
     \mu&^j_{k:k+1 | N}(z_k,z_{k+1}) = \myP^j_{k:k+1|N}(z_k,z_{k+1}) \nonumber \\
& \cdot \left(  (\mathbf{P}^i_{k:k+1 | k}(z_k))^{-1} \mu^i_{k:k+1|k}(z_k) -\gamma^\ell_{k+1}(z_{k+1})  \right), \\
\myP&^j_{k:k+1|N}(z_k,z_{k+1}) = \left((\myP^i_{k:k+1|k}(z_k))^{-1} + \begin{bmatrix}\mathbf{0} & \mathbf{0}\\ \mathbf{0} & \bar{\myL}_{k+1}^\ell(z_{k+1}) \end{bmatrix}\right)^{-1}, \\
\gamma&^\ell_{k+1}(z_{k+1}) = \begin{bmatrix} 0_{\text{nx} \times 1} \\ \bar{s}^\ell_{k+1}(z_{k+1})\end{bmatrix}, 
 \end{align}
\end{subequations}
and
\begin{subequations}
 \begin{align}
\mu&^i_{k:k+1|k} (z_k) = \begin{bmatrix} \mu_{k|k}^i(z_k) \\ \mathbf{A}_k(z_k)\mu^i_{k|k}(z_k) + b_k(z_k)\end{bmatrix}, \\
&\myP^i_{k:k+1|k}(z_k) \nonumber \\
&=  \begin{bmatrix} \myP^i_{k|k}(z_k) & \myP^i_{k|k}(z_k)\mathbf{A}^T_k(z_k) \\  \mathbf{A}_k(z_k)\myP^i_{k|k}(z_k) & \mathbf{A}_k(z_k)\myP^i_{k|k}(z_k)\mathbf{A}^T_k(z_k) +\mathbf{Q}_k(z_k) 
\end{bmatrix}.
 \end{align}
\end{subequations}
\end{lem}
\begin{pf}
We begin with
\begin{align}
p&(y_{k+1:N},x_{k+1},z_{k+1},x_k,z_k|y_{1:k}) \nonumber \\
&= p(x_{k+1},z_{k+1},x_k,z_k|y_{1:k},y_{k+1:N})p(y_{k+1:N}|y_{1:k}) \nonumber \\
&= p(x_{k+1},z_{k+1},x_k,z_k|y_{1:N})p(y_{k+1:N}|y_{1:k}). \label{eq:thin4556}
\end{align}
Therefore the RHS of \eqref{eq:thin4556} is equal to
\begin{align}
\label{eq:normconstjointsmoo44}
p&(x_{k+1},z_{k+1},x_k,z_k|y_{1:N})p(y_{k+1:N}|y_{1:k}) \nonumber \\
&=p(y_{k+1:N}|x_{k+1},z_{k+1},x_{k},z_{k},y_{1:k}) \nonumber \\
& \quad \cdot p(z_{k+1}|z_k,x_{k+1},x_{k},y_{1:k})   p(x_{k+1}|x_k,z_k,y_{1:k})p(x_k,z_k|y_{1:k}) ,\nonumber \\
&=p(y_{k+1:N}|x_{k+1},z_{k+1},x_{k},z_{k})p(z_{k+1}|z_k) \nonumber \\
& \quad \cdot p(x_{k+1}|x_k,z_k)  p(x_k,z_k|y_{1:k}).
\end{align}
And finally the LHS of \eqref{eq:thin4556} can be rearranged to give
\begin{align}
p&(x_{k+1},z_{k+1},x_k,z_k|y_{1:N}) 
=p(y_{k+1:N}|x_{k+1},z_{k+1},x_{k},z_{k}) \nonumber \\
&\cdot\frac {p(z_{k+1}|z_k)    p(x_{k+1}|x_k,z_k)  p(x_k,z_k|y_{1:k})}{p(y_{k+1:N}|y_{1:k})}.
\end{align}
Next, we can compute the denominator $p(y_{k+1:N}|y_{1:k})$  to be the normalising constant. Continuing from \eqref{eq:normconstjointsmoo44},
\begin{align}
&p(y_{k+1:N}|y_{1:k}) \nonumber \\
&= \sum_{z_{k+1},z_k} \int p(y_{k+1:N}|y_{1:k})p(x_{k+1},z_{k+1},x_k,z_k|y_{1:N}) \, dx_{k:k+1}\nonumber \\
&=\sum_{z_{k+1},z_k} \int p(y_{k+1:N}|x_{k+1},z_{k+1},x_{k},z_{k})p(z_{k+1}|z_k)  \nonumber \\
& \quad \cdot p(x_{k+1}|x_k,z_k)  p(x_k,z_k|y_{1:k}) \, dx_{k:k+1}.
\end{align}
%

Substituting the form of
  \begin{align}
p(z_{k+1},z_k) &= \mathbf{T}(z_{k+1},z_k), \nonumber \\
 p(x_{k+1}|x_k,z_k) &= \mathcal{N}(x_{k+1}|\mathbf{A}_k(z_k)x_k+b_k(z_k),\mathbf{Q}_k(z_k)), \nonumber \\
 p(x_k,z_k | y_{1:k}) &= \sum_{i=1}^\numFilt{k} w^i_{k | k}(z_k) \cdot \mathcal{N}  (x_k |  \mu^i_{k| k}(z_k),  \myP^i_{k | k}(z_k)  ),
  \end{align}
  and the properties of Normal distributions yields
  \begin{tiny}
\begin{align}
p&(x_{k+1}|x_k,z_k) p(z_{k+1}|z_k)  p(x_k,z_k | y_{1:k}) \nonumber \\
&= \sum_{i=1}^\numFilt{k} \mathbf{T}(z_{k+1},z_k) w^i_{k|k}(z_k) \mathcal{N}  (x_k |  \mu^i_{k| k}(z_k),  \myP^i_{k | k}(z_k)  ) \nonumber \\
& \cdot \mathcal{N}(x_{k+1}|\mathbf{A}_k(z_k)x_k+b_k(z_k),\mathbf{Q}_k(z_k)) \nonumber \\
& = \sum_{i=1}^\numFilt{k} w^i_{k+1|k}(z_k,z_{k+1})  \mathcal{N}(\chi_k | \mu^i_{k:k+1|k}(z_k),\myP^i_{k:k+1|k}(z_k) ),
\end{align}
\end{tiny}
  
where

\begin{align}
& w^i_{k+1 | k}(z_k,z_{k+1})=  \mathbf{T}(z_{k+1},z_k) w^i_{k | k}(z_k), \label{eq:topexprre43}\\
\mu&^i_{k:k+1|k} (z_k) = \begin{bmatrix} \mu_{k|k}^i(z_k) \\ \mathbf{A}_k(z_k)\mu^i_{k|k}(z_k) + b_k(z_k)\end{bmatrix}, \\
&\myP^i_{k:k+1|k}(z_k) \nonumber \\
&=  \begin{bmatrix} \myP^i_{k|k}(z_k) & \myP^i_{k|k}(z_k)\mathbf{A}^T_k(z_k) \\  \mathbf{A}_k(z_k)\myP^i_{k|k}(z_k) & \mathbf{A}_k(z_k)\myP^i_{k|k}(z_k)\mathbf{A}^T_k(z_k) +\mathbf{Q}_k(z_k) 
\end{bmatrix}.
\end{align}

Importantly, a square-root factor exists,
\begin{tiny}
\begin{align}
 (\myP^i_{k:k+1|k}(z_k))^{1/2} = \begin{bmatrix} (\mathbf{P}^i_{k|k}(z_k) )^{1/2} & (\mathbf{P}^i_{k|k}(z_k))^{1/2} \mathbf{A}_k^T(z_k)  \\ \mathbf{0} & \mathbf{Q}_k^{1/2}(z_k) \end{bmatrix}.
\end{align}
\end{tiny}

The other component required for joint-smoothing is the backward likelihood
\begin{align}
&p(y_{k+1:N} \mid x_{k+1},z_{k+1})\nonumber \\
&= \sum_{\ell =1}^{\numCorr{k+1}} \mathcal{L} \left ( x_{k+1} \, ;\, \bar{r}_{k+1}^\ell(z_{k+1}),\, \bar{s}_{k+1}^\ell(z_{k+1}),\, \bar{\myL}_{k+1}^\ell(z_{k+1}) \right ).
\end{align}

As the likelihood components are exponentials of quadratic functions
we can extend the functions to take in independent variables as arguments, 
\begin{align}
&x_{k+1}^T\myL x_{k+1} + 2 x^T_{k+1}s + r  = \chi_k^T  \begin{bmatrix}\mathbf{0} & \mathbf{0} \\ \mathbf{0}  & \myL \end{bmatrix} \chi_k +  2\chi_k^T  \begin{bmatrix}\vec{0}\\ s \end{bmatrix}+r.
\end{align}

This yields
\begin{align}
&p(y_{k+1:N} \mid x_{k+1},z_{k+1},x_{k},z_{k}) \\
&= \sum_{\ell =1}^{\numCorr{k+1}} \mathcal{L}  (\chi_k \, ;\, {r}_{k:k+1}^\ell(z_{k+1}),\, {s}_{k:k+1}^\ell(z_{k+1}),\, {\myL}_{k:k+1}^\ell(z_{k+1})  ),
\end{align}
where
\begin{align}
\label{eq:augementexp344}
{\myL}_{k:k+1|k+1}^\ell(z_{k+1}) &= \begin{bmatrix}\mathbf{0} & \mathbf{0} \\ \mathbf{0}  & \bar{\myL}_{k+1}^\ell(z_{k+1})  \end{bmatrix}, \nonumber \\
{s}_{k:k+1|k+1}^\ell(z_{k+1}) &= \begin{bmatrix}\vec{0}\\ \bar{s}_{k+1}^\ell(z_{k+1})  \end{bmatrix},  \ 
{r}_{k:k+1|k+1}^\ell(z_{k+1}) = \bar{r}_{k+1}^\ell(z_{k+1}).
\end{align}
We can now form
\begin{align}
&p(y_{k+1:N}|y_{1:k}) p(x_{k+1},z_{k+1},x_k,z_k|y_{1:N}) \nonumber \\
&=p(y_{k+1:N}|x_{k+1},z_{k+1},x_{k},z_{k})p(z_{k+1}|z_k) \nonumber \\
 &\quad \cdot p(x_{k+1}|x_k,z_k)  p(x_k,z_k|y_{1:k}) \nonumber \\
 &  =  \sum_{i=1}^\numFilt{k} w^i_{k+1|k}(z_k,z_{k+1})  \mathcal{N}\left(\chi_k | \mu^i_{k:k+1|k}(z_k),\myP^i_{k:k+1|k}(z_k) \right) \nonumber \\
& \cdot \sum_{\ell =1}^{\numCorr{k+1}} \mathcal{L} \left (\chi_k \, ;\, {r}_{k:k+1}^\ell(z_{k+1}),\, {s}_{k:k+1}^\ell(z_{k+1}),\, {\myL}_{k:k+1}^\ell(z_{k+1}) \right ).
\end{align}
By combining the sums into a single index $j$, Lemma~\ref{lem:smoothedmode} can be applied to give 
\begin{align}
\label{eq:subintomesmooth455}
&p(y_{k+1:N}|y_{1:k}) p(x_{k+1},z_{k+1},x_k,z_k|y_{1:N}) \nonumber \\
&= \sum_{j=1}^\numJSmooth{k} \tilde{w}^j_{k:k+1|N}(z_k,z_{k+1})\nonumber \\
&\cdot\mathcal{N}( \chi_k|\mu_{k:k+1|N}^j(z_k,z_{k+1}),\myP_{k:k+1|N}^j(z_k,z_{k+1})),
\end{align}
with the statistics
\begin{subequations}
\begin{align}
&\myP^j_{k:k+1|N}(z_k,z_{k+1}) = ((\myP_{k:k+1|k}^i(z_k))^{-1} + \myL^\ell_{k:k+1|k+1}(z_{k+1}))^{-1},\\
&{\mu}_{k:k+1|N}^j(z_k,z_{k+1}) = \myP_{k:k+1|N}^i(z_k) ((\myP^i_{k:k+1|k}(z_k))^{-1}\mu^i_{k:k+1|k}(z_k) \nonumber\\
& \quad - s^\ell_{k:k+1|k+1}(z_{k+1})),\\
%
\tilde{w}&_{k:k+1 \mid N}^j(z_k,z_{k+1}) = e^{\frac{1}{2}\beta^j(z_k,z_{k+1})} ,\\ 
        \beta&^j(z_k,z_{k+1}) =  2\ln(w^i_{k|k}(z_k)) + 2\ln(\mathbf{T}(z_{k+1},z_k))\nonumber \\
        &+ (\mu^j_{k:k+1|N}(z_k,z_{k+1}))^T(\myP_{k:k+1|N}^j(z_k,z_{k+1}))^{-1}\mu^j_{k:k+1|N}(z_k,z_{k+1}) \nonumber \\
    &\quad -  (\mu^i_{k:k+1 \mid k}(z_k))^T (\myP^i_{k:k+1 | k}(z_k))^{-1} \mu^i_{k:k+1 | k}(z_k)  \nonumber \\
    &\quad  -\bar{r}^\ell_{k+1}(z_{k+1}) +\ln|\myP_{k:k+1|N}^j(z_k,z_{k+1})| - \ln|\myP_{k:k+1|k}^i(z_k)|  .
\end{align}
\end{subequations}
Substituting \eqref{eq:augementexp344} and \eqref{eq:topexprre43} yields many of the required expressions.
Finally the normalising constant is computed as
\begin{align}
&p(y_{k+1:N}|y_{1:k}) =\sum_{z_k,z_{k+1}} \int p(y_{k+1:N}|x_{k+1},z_{k+1},x_{k},z_{k})p(z_{k+1}|z_k)  \nonumber \\
& \quad \cdot p(x_{k+1}|x_k,z_k)  p(x_k,z_k|y_{1:k}) \, dx_{k:k+1}  \nonumber \\
&=\sum_{z_k,z_{k+1}} \int   \sum_{j=1}^{\numJSmooth{k}} \tilde{w}^j_{k:k+1 | N}(z_k,z_{k+1}) \nonumber \\
  & \cdot \mathcal{N} \left (x_k\, ; \, \mu^j_{k:k+1
                              \mid N}(z_k,z_{k+1}), \, \myP^j_{k:k+1 \mid N}(z_k,z_{k+1})
                              \right ), \, dx_{k:k+1} \nonumber \\
&=\sum_{z_k,z_{k+1}}    \sum_{j=1}^{\numJSmooth{k}} \tilde{w}^j_{k:k+1 | N}(z_k,z_{k+1}) .
\end{align}
Substituting this into \eqref{eq:subintomesmooth455} and defining
\begin{align}
{w}^j_{k:k+1 | N}(z_k,z_{k+1})=\frac{\tilde{w}^j_{k:k+1 | N}(z_k,z_{k+1})}{\sum_{z_k,z_{k+1}}    \sum_{j=1}^{\numJSmooth{k}} \tilde{w}^j_{k:k+1 | N}(z_k,z_{k+1})}
\end{align}
yields
\begin{align}
& p(x_{k+1},z_{k+1},x_k,z_k|y_{1:N}) = \sum_{j=1}^\numJSmooth{k} {w}^j_{k:k+1|N}(z_k,z_{k+1})\nonumber \\
&\cdot\mathcal{N}\left( \chi_k\bigg|\mu_{k:k+1|N}^j(z_k,z_{k+1}),\myP_{k:k+1|N}^j(z_k,z_{k+1})\right).
\end{align}
%
\end{pf}

\begin{lem}
\label{lem:smoothedmode}
Let ${\bar{\mathbf{L}}}$, $\bar{s}$, and $\bar{r}$ be the information matrix, information vector and information scalar respectively, the sufficient statistics for a likelihood mode in the BIF.
Additionally, let $\myP$, $\mu$, and $w$ be the covariance matrix, mean and weight of a Gaussian mode. Then the statistics of the combined smoothed mode $\{\bar{\myP}, \bar{\mu}, \bar{w} \}$ can be computed as
\begin{align} \bar{w} \mathcal{N}(x|\bar{\mu},\bar{\myP}) = w\mathcal{N} (x|\mu,\myP) \mathcal{L}(x|r,s,\myL),\end{align}
where
\begin{subequations}
\begin{align}
\bar{\myP} &= (\myP^{-1} + \myL)^{-1}, \ \bar{\mu} = \bar{\myP}(\myP^{-1}\mu - s),\ \bar{w} = e^{\frac{1}{2}\beta}, \nonumber \\
\beta &= 2\ln(w)+ \ln|2\pi \bar{\myP}| - \ln|2\pi \myP| + \bar{\mu}^T\bar{\myP}^{-1}\bar{\mu} - \mu^T \myP^{-1}\mu - r.
\end{align}
\end{subequations}
\end{lem}
\begin{pf}
We begin with
\begin{align}
&w\mathcal{N} (x|\mu,\myP) \mathcal{L}(x|r,s,\myL) \nonumber \\
&=\frac{we^{-\frac{1}{2}(\mu^T\myP^{-1}\mu + r)}}{{|2\pi \myP|}^{\frac{1}{2}}}e^{-\frac{1}{2}\left(x^T(\myP^{-1}+\myL)x-2x^T(\myP^{-1}\mu-s)\right)}  \nonumber \\
&=\frac{w{|2\pi \bar{\myP}|}^{\frac{1}{2}}e^{-\frac{1}{2}(\mu^T\myP^{-1}\mu + r -\bar{\mu}^T\bar{\myP}^{-1}\bar{\mu})}}{{|2\pi \myP|}^{\frac{1}{2}}{|2\pi \bar{\myP}|}^{\frac{1}{2}}}   e^{-\frac{1}{2}(x^T\bar{\myP}^{-1}x-2x^T\bar{\myP}^{-1}\bar{\mu} +\bar{\mu}^T\bar{\myP}^{-1}\bar{\mu})}.
\end{align}
where $\bar{\myP}^{-1} = \myP^{-1} + \myL$, and $\bar{\mu} = \bar{\myP}(\myP^{-1}\mu - s)$.
Substitute for $\mathcal{N}(x|\bar{\mu},\bar{\myP})$ yields
\begin{align}
&w\mathcal{N} (x|\mu,\myP) \mathcal{L}(x|r,s,\myL) =\frac{w{|2\pi \bar{\myP}|}^{\frac{1}{2}}e^{\frac{1}{2}\bar{\beta}}}{{|2\pi \myP|}^{\frac{1}{2}}} \mathcal{N}(x|\bar{\mu},\bar{\myP})
=e^{\frac{1}{2}{\beta}}\mathcal{N}(x|\bar{\mu},\bar{\myP}),
\end{align}
where
\begin{align}
\beta = 2\ln(w)+ \ln|2\pi \bar{\myP}| - \ln|2\pi \myP| + \bar{\mu}^T\bar{\myP}^{-1}\bar{\mu} - \mu^T \myP^{-1}\mu - r.
\end{align}
\end{pf}
\end{small}
\clearpage
\section{EM Lemmata}
\begin{small}
For the following Lemma, we use the shorthand
\begin{align}
\zeta_k = \begin{bmatrix}x_k \\ u_k\end{bmatrix}, \quad \eta_k = \begin{bmatrix}y_k \\ x_{k+1}\end{bmatrix}
\end{align}

\subsection{Proof of Lemma~\ref{lem:JMLS_EM_soln345}}
We begin with the Q function,
\begin{align}
\label{eq:Qfunctappendix45}
\mathcal{Q}(\theta,\theta') \triangleq  \sum_{z_{1:N+1}}\int \ln&\left(\pt \left(x_{1:N+1},z_{1:N+1},y_{1:N}\right)\right) \nonumber \\ &
\cdot \ptp(x_{1:N+1},z_{1:N+1}|y_{1:N}) \, dx_{1:N+1},
\end{align}
where
\begin{align}
\pt \left(x_{1:N+1},z_{1:N+1},y_{1:N}\right) &= \pt(x_1,z_1) \prod_{k=1}^{N} p\left(x_{k+1},z_{k+1},y_k|x_k,z_k\right),\\
\pt\left(x_{k+1},z_{k+1},y_k|x_k,z_k\right) &= \pt\left(x_{k+1},y_k|x_k,z_k\right)\pt(z_{k+1}|z_k,\hcancel{x_k}).
\end{align}
Therefore
\begin{align}
\label{eq:logjoint457}
\ln&\left(\pt \left(x_{1:N+1},z_{1:N+1},y_{1:N}\right)\right)=\ln(\pt(x_1,z_1)) \nonumber \\
& + \sum_{k=1}^{N} \ln(\pt(x_{k+1},y_k|x_k,z_k)) + \sum_{k=1}^{N} \ln(\pt(z_{k+1}|z_k)).
\end{align}

Substituting \eqref{eq:logjoint457} into \eqref{eq:Qfunctappendix45}, integrating over independent continuous variables, and applying Lemma~\ref{lem:reducedZtraj} yields
\begin{tiny}
\begin{align}&Q(\theta,\theta') = m^{N-1}\sum_{z_{1}}\int   \ln[{\pt}_1(x_1,z_1)]    \ptp(x_{1},z_{1}|y_{1:N}) \, dx_{1} \nonumber \\
&+ m^{N-2} \sum_{k=1}^{N}  \int   \sum_{z_{k:k+1}}  \ln[{\pt}_T\left(z_{k+1}|z_k \right)]     \ptp(x_{k:k+1},z_{k:k+1}|y_{1:N}) \, dx_{k:k+1} \nonumber \\
&+ m^{N-2} \sum_{k=1}^{N}  \int   \sum_{z_{k:k+1}}  \ln[{\pt}_L\left(x_{k+1},y_k|x_k,z_k \right)]    \ptp(x_{k:k+1},z_{k:k+1}|y_{1:N}) \, dx_{k:k+1}.
\end{align}
\end{tiny}
Note that $\theta$ is now separated into different components, as $\theta = \{\theta_1,\theta_L,\theta_T \}$, each of these terms can be independently maximized for the optimal parameter set.

\textbf{Optimising for ML prior distribution}\\
Using the Kullback--Leibler divergence of a hybrid distribution we arrive at the inequality
\begin{align}
\sum_z \int \ln (p(x,z)) p(x,z) \, dx \geq \sum_z \int  \ln (q(x,z)) p(x,z) \, dx .
\end{align}
Therefore the maximiser of the cost function
\begin{align} c_1 \sum_{z_{1}}\int  \ln\left( p_{\theta_1}(x_1,z_1)  \right)  \ptp(x_{1},z_{1}|y_{1:N}) \, dx_{1},\end{align}
is with the prior distribution
\begin{align}p_{\theta_1}(x_1,z_1)  =  \ptp(x_{1},z_{1}|y_{1:N}). \end{align}

\textbf{Optimising for ML transition matrix}\\
Note that not all terms in the sum will contain the transition probability parameter we are interested in, and that the sum over discrete trajectories adjust accordingly.
Taking the derivative
\begin{tiny}
\begin{align}
&\frac{\partial Q(\theta,\theta')}{\partial \mathbf{T}(a,b) }=0  = m^{N-2} \sum_{k=1}^{N} \sum_{j=1}^{\numJSmooth{k}(b,a)} w^j_{k:k+1|N}(b,a) \bigg( \frac{ \partial }{\partial \mathbf{T}(a,b) } \ln\left(\mathbf{T}(a,b)\right)     \bigg) \nonumber \\
&\quad -  m^{N-2} \sum_{k=1}^{N}\sum_{z_{k+1}} \sum_{j=1}^{\numJSmooth{k}(b,z_{k+1})} w^j_{k:k+1|N}(b,z_{k+1})   \bigg( \frac{ \partial }{\partial \mathbf{T}(a,b) }  \ln\left(\sum_{i=1}^{m}\mathbf{T}(i,b)\right)      \bigg),
\end{align}
\begin{align}
0=&  \sum_{k=1}^{N} \sum_{j=1}^{\numJSmooth{k}(b,a)} \frac{w^j_{k:k+1|N}(b,a)}{\mathbf{T}(a,b)}   \nonumber \\
&- \sum_{k=1}^{N}\sum_{z_{k+1}} \sum_{j=1}^{\numJSmooth{k}(b,z_{k+1})} w^j_{k:k+1|N}(b,z_{k+1}) \left(  \frac{ 1 }{\sum_{i=1}^{m}\mathbf{T}(i,b) }\cdot 1 \right),
\end{align}
\end{tiny}
where $\sum_{i=1}^{m}\mathbf{T}(i,b) =1 \quad \forall \, b=1,\dots, m$
\begin{tiny}
\begin{align}
& \frac{1}{\mathbf{T}(a,b)} \sum_{k=1}^{N} \sum_{j=1}^{\numJSmooth{k}(b,a)} w^j_{k:k+1|N}(b,a) =
 \sum_{k=1}^{N}\sum_{z_{k+1}} \sum_{j=1}^{\numJSmooth{k}(b,z_{k+1})} w^j_{k:k+1|N}(b,z_{k+1}).
\end{align}
\end{tiny}
Therefore, by defining $c_m(b)$, we yield
\begin{align}
c_m(b) &= \sum_{k=1}^{N}\sum_{z_{k+1}} \sum_{j=1}^{\numJSmooth{k}(b,z_{k+1})} w^j_{k:k+1|N}(b,z_{k+1}), \\
\mathbf{T}(a,b) &=\frac{1}{c_m(b)}\sum_{k=1}^{N} \sum_{j=1}^{\numJSmooth{k}(b,a)} w^j_{k:k+1|N}(b,a).
\end{align}

\textbf{Optimising for ML linear parameters}\\
Taking the derivative wrt the Q function yields
\begin{align}
&\frac{\boldsymbol{\Pi}(b)\partial \mathcal{Q}(\theta,\theta')}{\partial\boldsymbol{\Gamma}(b)} =0  =\boldsymbol{\Pi}(b)\frac{\partial }{\partial\boldsymbol{\Gamma}(b)} m^{N-2}\nonumber \\
& \cdot \sum_{k=1}^{N}\sum_{z_{k+1}} \sum_{j=1}^{\numJSmooth{k}(b,z_{k+1})} w^j_{k:k+1|N}(b,z_{k+1})  \expected_{k}^{b,z_{k+1},j}[\ln(  \N ( \eta_k | \boldsymbol{\Gamma}(b)\zeta_k,\boldsymbol{\Pi}(b) ))] .
\end{align}
Thus
\begin{align}
&0= m^{N-2} \sum_{k=1}^{N}\sum_{z_{k+1}} \sum_{j=1}^{\numJSmooth{k}(b,z_{k+1})} w^j_{k:k+1|N}(b,z_{k+1})  \expected_{k}^{b,z_{k+1},j}\left[\eta_k \zeta_k^T \right]
\nonumber \\
& -m^{N-2} \boldsymbol{\Gamma}(b) \sum_{k=1}^{N}\sum_{z_{k+1}} \sum_{j=1}^{\numJSmooth{k}(b,z_{k+1})} w^j_{k:k+1|N}(b,z_{k+1}) \expected_{k}^{b,z_{k+1},j}\left[\zeta_k\zeta_k^T\right].
\end{align}


Therefore
\begin{subequations}
\begin{align}
\boldsymbol{\Gamma}(b) &= \boldsymbol{\Psi}(b) \boldsymbol{\Sigma}^{-1}(b) \label{eq:gammaEMest221}, 
\end{align}
where
\begin{align}
&\boldsymbol{\Sigma}(b) = \sum_{k=1}^{N}\sum_{z_{k+1}} \sum_{j=1}^{\numJSmooth{k}(b,z_{k+1})} w^j_{k:k+1|N}(b,z_{k+1}) \expected_{k}^{b,z_{k+1},j}\left[\zeta_k\zeta_k^T\right] ,
\\
&\boldsymbol{\Psi}(b) =\sum_{k=1}^{N}\sum_{z_{k+1}} \sum_{j=1}^{\numJSmooth{k}(b,z_{k+1})} w^j_{k:k+1|N}(b,z_{k+1}) \expected_{k}^{b,z_{k+1},j}\left[\eta_k \zeta_k^T \right].
\end{align}
\end{subequations}

Now calculating the derivative w.r.t. covariance parameters
%
\begin{tiny}
\begin{align}
&2\boldsymbol{\Pi}(b)   \frac{\partial\mathcal{Q}(\theta,\theta')}{\partial \boldsymbol{\Pi}(b)}\boldsymbol{\Pi}(b) =0 \nonumber \\
=&2\boldsymbol{\Pi}(b) \bigg( \frac{\partial}{\partial \boldsymbol{\Pi}(b)} \sum_{k=1}^{N}\sum_{z_{k+1}} \sum_{j=1}^{\numJSmooth{k}(b,z_{k+1})} w^j_{k:k+1|N}(b,z_{k+1}) \nonumber \\
&\cdot \expected_{k}^{b,z_{k+1},j}\left[\ln\left(  \N \left( \eta_k \bigg| \boldsymbol{\Gamma}(b)\zeta_k,\boldsymbol{\Pi}(b) \right)\right)\right] \bigg) \boldsymbol{\Pi}(b) \nonumber \\
&0= \sum_{k=1}^{N}\sum_{z_{k+1}} \sum_{j=1}^{\numJSmooth{k}(b,z_{k+1})} w^j_{k:k+1|N}(b,z_{k+1})  \expected_{k}^{b,z_{k+1},j}\left[ \eta_k\eta_k^T \right]  \nonumber \\
&- \sum_{k=1}^{N}\sum_{z_{k+1}} \sum_{j=1}^{\numJSmooth{k}(b,z_{k+1})} w^j_{k:k+1|N}(b,z_{k+1}) \expected_{k}^{b,z_{k+1},j}\left[ \eta_k\zeta_k^T  \right] \boldsymbol{\Gamma}^T(b) \nonumber \\
&- \sum_{k=1}^{N}\sum_{z_{k+1}} \sum_{j=1}^{\numJSmooth{k}(b,z_{k+1})} w^j_{k:k+1|N}(b,z_{k+1}) \boldsymbol{\Gamma}(b)  \expected_{k}^{b,z_{k+1},j}\left[ \zeta_k\eta_k^T \right]  \nonumber \\
&+ \sum_{k=1}^{N}\sum_{z_{k+1}} \sum_{j=1}^{\numJSmooth{k}(b,z_{k+1})} w^j_{k:k+1|N}(b,z_{k+1}) \boldsymbol{\Gamma}(b) \expected_{k}^{b,z_{k+1},j}\left[\zeta_k\zeta_k^T  \right] \boldsymbol{\Gamma}^T(b)  \nonumber \\
&- \sum_{k=1}^{N}\sum_{z_{k+1}} \sum_{j=1}^{\numJSmooth{k}(b,z_{k+1})} w^j_{k:k+1|N}(b,z_{k+1}) \boldsymbol{\Pi}(b).
\end{align}
\end{tiny}


Therefore
\begin{tiny}
\begin{align}
%
%
 \boldsymbol{\Pi}(b) &= \frac{1}{c_m(b)} \Big( \boldsymbol{\Phi}(b) -\boldsymbol{\Psi}(b) \boldsymbol{\Gamma}^T (b) - \boldsymbol{\Gamma} (b) \boldsymbol{\Psi}^T(b) + \boldsymbol{\Gamma} (b) \boldsymbol{\Sigma} (b) \boldsymbol{\Gamma}^T (b) \Big)  , 
 \end{align}
 \end{tiny}
where
 \begin{align}
%
%
\boldsymbol{\Phi}(b) &= \sum_{k=1}^{N}\sum_{z_{k+1}} \sum_{j=1}^{\numJSmooth{k}(b,z_{k+1})} w_{k:k+1|N}^j (b,z_{k+1})   \expected_{k}^{b,z_{k+1},j} \left[ \eta_k \eta_k^T \right] .
\end{align} 

\begin{lem}
\label{lem:reducedZtraj}
Consider a function of part of the discrete Markov chain within a sum of all possible state trajectories of the entire Markov chain, where the number of models available at each time step is the constant $m$, i.e.
\begin{align}\sum_{z_{1:N}} f(z_{k:k+n}) = \sum_{z_1=1}^{m}\dots \sum_{z_{N}=1}^{m}f(z_{k:k+n}),\end{align}
then 
\begin{align}\sum_{z_{1:N}} f(z_{k:k+n}) =m^{N-n-1} \sum_{z_{k:k+n}} f(z_{k:k+n}) .\end{align}
\end{lem}
\begin{pf}
Rearranging the original expression yields
\begin{tiny}
\begin{align}
&\sum_{z_{1:N}} f(z_{k:k+n})=\sum_{z_{1:N}} f(z_{k:k+n}) \times 1  = \sum_{z_k=1}^{m}\dots\sum_{z_{k+n}=1}^{m} f(z_{k:k+n})  \nonumber \\
&\cdot \sum_{z_1=1}^{m}\sum_{z_2=1}^{m}\dots \sum_{z_{k-1}=1}^m \sum_{z_{k+n+1}=1}^m \dots \sum_{z_{N-1}=1}^{m}\sum_{z_{N}=1}^{m} 1 \nonumber \\ 
&=m^{N-n-1}\sum_{z_{k:k+n}} f(z_{k:k+n}).\end{align}
\end{tiny}
\end{pf}
\end{small}

\subsection{Proof of Lemma~\ref{lem:sqrtmassExpect45768}}
Suppose that the square-root factor of the following expectation exists
\begin{tiny}
\begin{align}
\expected_{k}\left[\begin{bmatrix}x_k\\x_{k+1}\\u_k\\y_k\end{bmatrix}\begin{bmatrix}x_k\\x_{k+1}\\u_k\\y_k\end{bmatrix}^T\right]^{\frac{1}{2}} = \begin{bmatrix} M_{11}&M_{12}\\0 & M_{22}\end{bmatrix},
\end{align}
\end{tiny}
where $M_{11}$ and $M_{22}$ are UT matrices.
By equating parts of the joint expectation, we can form following required equalities
\begin{tiny}
\begin{align}
&M_{11}^TM_{11} \label{eq:M11eq}= \expected_{k}  \left[\begin{bmatrix}x_k\\x_{k+1}\end{bmatrix}\begin{bmatrix}x_k\\x_{k+1}\end{bmatrix}^T\right]  
= \mu_{k:k+1|N}\mu^T_{k:k+1|N}+\myP_{k:k+1|N}, \\
&\label{eq:refmelaterM12}M_{11}^TM_{12} = \expected_{k} \left[\begin{bmatrix}x_k\\x_{k+1}\end{bmatrix}\begin{bmatrix}u_k\\y_k\end{bmatrix}^T\right] 
 =\mu_{k:k+1|N}\begin{bmatrix} u^T_k & y^T_k\end{bmatrix},\\
&\label{eq:lastsqreexpejoin54}M_{12}^TM_{12}+M_{22}^TM_{22}=\expected_{k} \left[\begin{bmatrix}u_k\\y_k\end{bmatrix}\begin{bmatrix}u_k\\y_k\end{bmatrix}^T\right]
=\begin{bmatrix} u_k \\ y_k\end{bmatrix}\begin{bmatrix} u^T_k & y^T_k\end{bmatrix}.
\end{align}
\end{tiny}
From \eqref{eq:M11eq}, applying well-known QR pattern yields
\begin{align}
M_{11} &= (\myP_{k:k+1|N}+\mu_{k:k+1|N}\mu^T_{k:k+1|N})^{1/2}   =\mQb\begin{bmatrix} \myP^{1/2}_{k:k+1|N} \\  \mu^T_{k:k+1|N}\end{bmatrix}. \end{align}

Since $M_{11}$ is guaranteed to be invertible, we can rearrange \eqref{eq:refmelaterM12} and define
\begin{align}
\lambda \triangleq M_{11}^{-T}\mu_{k:k+1|N},
\end{align}
 to yield
\begin{align}
\label{eq:M12joingmassexpect443}
M_{12}=\lambda\begin{bmatrix} u^T_k & y^T_k\end{bmatrix}.
\end{align}
Finally by substituting \eqref{eq:M12joingmassexpect443} into \eqref{eq:lastsqreexpejoin54}, then noting $\lambda^T\lambda$ as a scalar quantity yields
\begin{align}
&\begin{bmatrix} u_k \\ y_k\end{bmatrix}\lambda^T\lambda\begin{bmatrix} u^T_k & y^T_k\end{bmatrix}+M_{22}^TM_{22}
= \begin{bmatrix} u_k \\ y_k\end{bmatrix}\begin{bmatrix} u^T_k & y^T_k\end{bmatrix},\\
\therefore &M_{22}^TM_{22} =\begin{bmatrix} u_k & y_k\end{bmatrix}(1-\lambda^T\lambda)\begin{bmatrix} u^T_k & y^T_k\end{bmatrix}.
\end{align}
The most obvious choice for a square-root factor is
\begin{align}M_{22} = \sqrt{1-\lambda^T\lambda}\begin{bmatrix} u^T_k & y^T_k\end{bmatrix}.\end{align}

This relies on $\lambda^T\lambda \leq 1$ which, after substituting for $\lambda$ requires
$$\mu^T_{k:k+1|N}(\myP_{k:k+1|N}+\mu_{k:k+1|N}\mu^T_{k:k+1|N})^{-1}\mu_{k:k+1|N}\leq 1,$$
which is true by application of the Sherman-Morrison formula, and noting that since $\myP$ is positive-definite $\mu^T\myP^{-1}\mu >0$.
\subsection{Proof of Lemma~\ref{sec:M-stepqrt346}}
We begin with the equating for calculating the covariance for the $z_\text{th}$ model
\begin{tiny}
\begin{align}
\label{eq:54765476576432409}
\boldsymbol{\Pi}(z) &=  \frac{1}{c_m(z)} \Big( \boldsymbol{\Phi}(z) -\boldsymbol{\Psi}(z) \boldsymbol{\Gamma}^T (z) - \boldsymbol{\Gamma} (z) \boldsymbol{\Psi}^T(z)  + \boldsymbol{\Gamma} (z) \boldsymbol{\Sigma} (z) \boldsymbol{\Gamma}^T (z) \Big) \nonumber \\
&=\frac{1}{c_m(z)} \begin{bmatrix}\mathbf{I} \\ -\boldsymbol{\Gamma}^T(z) \end{bmatrix}^T \begin{bmatrix}\boldsymbol{\Phi}(z) & \boldsymbol{\Psi}(z) \\ \boldsymbol{\Psi}^T(z) & \boldsymbol{\Sigma}(z)  \end{bmatrix} \begin{bmatrix}\mathbf{I} \\ -\boldsymbol{\Gamma}^T(z) \end{bmatrix}.
\end{align}
\end{tiny}
The matrix containing the expectation of sufficient statistics can also be written as
\begin{align}
&\begin{bmatrix}\boldsymbol{\Phi}(z) & \boldsymbol{\Psi}(z) \\ \boldsymbol{\Psi}^T(z) & \boldsymbol{\Sigma}(z)  \end{bmatrix}  
&=\begin{bmatrix} T_2\\ T_1 \end{bmatrix} \mathbf{M}(z) \begin{bmatrix} T_2\\ T_1 \end{bmatrix}^T, \label{eq:expectM5474378}
\end{align}
where 
\begin{align}
T_1 &= \begin{bmatrix}\mathbf{I}_{n_x} & \mathbf{0}_{n_x}& \mathbf{0}_{n_u}&\mathbf{0}_{n_y} \\
\mathbf{0}_{n_x} & \mathbf{0}_{n_x}& \mathbf{I}_{n_u}&\mathbf{0}_{n_y},
\end{bmatrix},\ 
T_2 = \begin{bmatrix}\mathbf{0}_{n_x} & \mathbf{0}_{n_x}& \mathbf{0}_{n_u}&\mathbf{I}_{n_y} \\
\mathbf{0}_{n_x} & \mathbf{I}_{n_x}& \mathbf{0}_{n_u}&\mathbf{0}_{n_y}
\end{bmatrix},
\end{align}
\begin{tiny}
\begin{align}
\mathbf{M}(z) &=   \sum_{k=1}^{N}  \sum_{j=1}^{m}    \sum_{\ell=1}^{\numJSmooth{k}}w_{k:k+1|N}^\ell (\be,j)  \expected_{k}^{\be,j,\ell} \left[\begin{bmatrix}x_k\\x_{k+1}\\u_k\\y_k\end{bmatrix}\begin{bmatrix}x_k\\x_{k+1}\\u_k\\y_k\end{bmatrix}^T\right].
\end{align}
\end{tiny}

Equating parts of \eqref{eq:expectM5474378} yields the equations for calculating the expectation of sufficient statistics when $\mathbf{M}^{1/2}(z)$ is available,
\begin{align}
\boldsymbol{\Sigma}(z) &= T_1 \left(\mathbf{M}^{1/2}(z)\right)^T\mathbf{M}^{1/2}(z)T_1^T, \\
\boldsymbol{\Phi}(z) &= T_2 \left(\mathbf{M}^{1/2}(z)\right)^T\mathbf{M}^{1/2}(z)T_2^T, \\
\boldsymbol{\Psi}(z) &= T_2 \left(\mathbf{M}^{1/2}(z)\right)^T\mathbf{M}^{1/2}(z)T_1^T, 
\end{align}
as $\mathbf{M}(z) = \left(\mathbf{M}^{1/2}(z)\right)^T\mathbf{M}^{1/2}(z)$.
Substituting \eqref{eq:expectM5474378} into \eqref{eq:54765476576432409} yields
\begin{align}
\boldsymbol{\Pi}(z) &= &\frac{1}{c_m(z)} \begin{bmatrix}\mathbf{I} \\ -\boldsymbol{\Gamma}^T(z) \end{bmatrix}^T 
\begin{bmatrix} T_2\\ T_1 \end{bmatrix} \mathbf{M}(z) \begin{bmatrix} T_2\\ T_1 \end{bmatrix}^T
 \begin{bmatrix}\mathbf{I} \\ -\boldsymbol{\Gamma}^T(z) \end{bmatrix}.
\end{align}
Taking advantage of the form of the equation yields
\begin{align} 
\boldsymbol{\Pi}^{1/2}(z) &= \frac{1}{\sqrt{c_m(z)}} \mathbf{M}^{1/2}(z) \begin{bmatrix} T_2\\ T_1 \end{bmatrix}^T
 \begin{bmatrix}\mathbf{I} \\ -\boldsymbol{\Gamma}^T(z) \end{bmatrix} \nonumber \\
&=\frac{1}{\sqrt{c_m(z)}} \mathbf{M}^{1/2}(z)  \begin{bmatrix} -\mathbf{C}^T(\be) & -\mathbf{A}^T(\be) \\ \mathbf{0}_{n_x\times n_y} & \mathbf{I}_{n_x} \\ -\mathbf{D}^T(\be)& -\mathbf{B}^T(\be) \\ \mathbf{I}_{n_y} & \mathbf{0}_{n_y \times n_x} \end{bmatrix}.
\end{align}
\subsection{Proof Lemma~\ref{lem:samemodelparaminit}}

We begin with examining input output behavior of the function from \cite{generalJMLSpaperBalenzuela}, which is used to compute the JMLS filtered distribution 
\begin{align}
&p(x_{k},z_{k}|y_{1:k}) \nonumber \\
&= f_\text{correct}(p(x_k,z_k|y_{1:k-1}),u_k,y_k, \boldsymbol{\Gamma}(z_k),\boldsymbol{\Pi}(z_k)).
\end{align}
As we have assumed that each parameter set is identical, the dependence on $z_k$ can be removed for the parameters $ \boldsymbol{\Gamma}(z_k)$ and $\boldsymbol{\Pi}(z_k)$.
Furthermore, if $p(x_k,z_k|y_{1:k-1})$ is not a function of $z_k$, we can conclude that $p(x_{k},z_{k}|y_{1:k})$ is not a function of $z_k$ as
\begin{align}
&p(x_{k},z_{k}|y_{1:k}) = \bar{f}_\text{correct}(p(x_k|y_{1:k-1}),u_k,y_k, \boldsymbol{\Gamma},\boldsymbol{\Pi}).
\end{align}
This will be proven shortly, however we have asserted this to be true for $k=1$, in the choice of prior.
Note the bar appearing over the function, which caters to the change in input distribution, where the substitute distribution $p(x_k|y_{1:k-1})$, this acknowledges that $p(x_k,z_k|y_{1:k-1})$ is indeed not a function of $z_k$, and all valid $z_k$ values are equally probable.

Using similar logic, with the asserted transition matrix from the Lemma 
\begin{align}&\mathbf{T}(z_{k+1},z_k)=\frac{1}{m} \quad \forall z_{k}=1,\dots,m \quad \forall z_{k+1}=1,\dots,m,\end{align}
the function prediction distribution can be rewritten
\begin{align}
&p(x_{k+1},z_{k+1}|y_{1:k}) \nonumber \\
&= f_\text{pred.}(p(x_k,z_k|y_{1:k}),u_k, \boldsymbol{\Gamma}(z_k),\boldsymbol{\Pi}(z_k),\textbf{T}(z_{k+1},z_k)) \nonumber \\
&= \bar{f}_\text{pred.}(p(x_k|y_{1:k}),u_k, \boldsymbol{\Gamma},\boldsymbol{\Pi},\frac{1}{m}),
\end{align}
which completes the filtering recursion, as we have just proved that the prediction is independent of $z_k$ variable.

The same logic is then applied to the backwards information filter operation
\begin{subequations}
\begin{align}
&p(y_N|x_N,z_N)  = f_\text{BIF init}(y_N,\boldsymbol{\Gamma}(z_N),\boldsymbol{\Pi}(z_N)) \nonumber \\
&= {f}_\text{BIF init}(y_N,\boldsymbol{\Gamma},\boldsymbol{\Pi}),\\
&p(y_{k+1:N}|x_k,z_k) \nonumber \\
&= f_\text{BIF bwd}(p(y_{k+1:N}|x_{k+1},z_{k+1}),u_k,\boldsymbol{\Gamma}(z_k),\boldsymbol{\Pi}(z_k), \nonumber \\
& \quad\textbf{T}(z_{k+1},z_k) \forall z_{k+1}=1,\dots,m) \nonumber \\
&= \bar{f}_\text{BIF bwd}(p(y_{k+1:N}|x_{k+1}),u_k,\boldsymbol{\Gamma},\boldsymbol{\Pi},\frac{1}{m}),\\
&p(y_k|x_k,z_k)  = f_\text{BIF corr}(p(y_{k+1:N}|x_k,z_k),u_k,y_k, \boldsymbol{\Gamma}(z_k),\boldsymbol{\Pi}(z_k)) \nonumber\\
&= \bar{f}_\text{BIF corr}(p(y_{k+1:N}|x_k),u_k,y_k, \boldsymbol{\Gamma},\boldsymbol{\Pi}),
\end{align}
\end{subequations}
and therefore the likelihood from the backwards information filter is invariant to the discrete variable $z$.
Extending this logic further to the joint-smoothed distribution
\begin{subequations}
\begin{align}
&p(x_{k+1},z_{k+1},x_k,z_k|y_{1:N}) \nonumber \\
&= f_\text{smooth}(p(y_{k+1}|x_{k+1},z_{k+1}),p(x_{k},z_{k}|y_{1:k}),u_k, \nonumber \\
&\quad \quad \boldsymbol{\Gamma}(z_k),\boldsymbol{\Pi}(z_k),\textbf{T}(z_{k+1},z_k)) \nonumber \\
&= \bar{f}_\text{smooth}(p(y_{k+1}|x_{k+1}),p(x_{k}|y_{1:k}),u_k,  \boldsymbol{\Gamma},\boldsymbol{\Pi},\frac{1}{m}).
\end{align}
\end{subequations}
The expectation of sufficient statistics 
\begin{align} &\{ c_m(z_k), \boldsymbol{\Sigma}(z_k), \boldsymbol{\Phi}(z_k), \boldsymbol{\Psi}(z_k)  \} \nonumber \\
& = f_\text{E-STEP}(p(x_{k+1},z_{k+1},x_k,z_k|y_{1:N})  \forall k=1,\dots,N ) \nonumber \\
& = \bar{f}_\text{E-STEP}(p(x_{k+1},x_k|y_{1:N})  \forall k=1,\dots,N )
\end{align}
As the joint-smoothed distribution is invariant to the variables $z_{k}$ and $z_{k+1}$,
$c_m(z_k)=c_m$, $\boldsymbol{\Sigma}(z_k)=\boldsymbol{\Sigma}$, $\boldsymbol{\Phi}(z_k)=\boldsymbol{\Phi}$, $\boldsymbol{\Psi}(z_k)=\boldsymbol{\Psi}$.

Finally the new parameter estimates
\begin{align}
&\{ \boldsymbol{\Gamma}(z_k), \boldsymbol{\Pi}(z_k) \} 
= f_{\text{M-step }\boldsymbol{\Pi},\boldsymbol{\Gamma}}(c_m(z_k), \boldsymbol{\Sigma}(z_k), \boldsymbol{\Phi}(z_k), \boldsymbol{\Psi}(z_k)) 
\nonumber \\
&= {f}_{\text{M-step }\boldsymbol{\Pi},\boldsymbol{\Gamma}}(c_m, \boldsymbol{\Sigma}, \boldsymbol{\Phi}, \boldsymbol{\Psi}) 
\end{align}
Therefore $\boldsymbol{\Gamma}(z_k)=\boldsymbol{\Gamma}$ and $\boldsymbol{\Pi}(z_k)=\boldsymbol{\Pi}$.
Finally, the transition probabilities
\begin{align}
&\mathbf{T}(z_{k+1},z_k) = f_\text{M-step \textbf{T}}(p(x_{k+1},z_{k+1},x_k,z_k|y_{1:N}), c_m(z_k)) \nonumber \\
&=\bar{f}_\text{M-step \textbf{T}}(p(x_{k+1},x_k|y_{1:N}), c_m)
\end{align}
as $\mathbf{T}(z_{k+1},z_k)$ is invariant to the variables $z_{k+1}$ and $z_k$, and \mbox{$\sum_{i=1}^m \textbf{T}(i,j)=1$}, then
\begin{align}\mathbf{T}(z_{k+1},z_k) =\frac{1}{m}.\end{align}
Therefore, the estimated system parameter set has satisfied the original conditions for this lemma, and subsequent iterations will also satisfy this condition, and produce model estimates with duplicate parameter sets.

\clearpage
\bibliographystyle{plain}        
\bibliography{autosam}           

\end{document}